\documentclass[12pt]{article} 
\usepackage[margin=1in]{geometry}
\usepackage[sectionbib]{natbib}
\usepackage[dvipsnames]{xcolor}
\usepackage{amsfonts}
\usepackage{multirow}
\usepackage{booktabs}
\usepackage{mathtools}
\usepackage{hyperref}
\usepackage{amssymb}
\usepackage{amsmath}
\usepackage{amsthm}

\hypersetup{
  colorlinks,
  linkcolor={red!50!black},
  citecolor={blue},
  urlcolor={blue!80!black}
}

\newcommand{\bo}{\boldsymbol }

\newcommand{\diag}{\text{\normalfont{diag}}}

\newcommand{\A}{\mathbf A}

\newcommand{\ba}{\boldsymbol a}

\newcommand{\Q}{\mathbf Q}

\newcommand{\bw}{\boldsymbol \omega}

\newcommand{\X}{\mathbf X}
\newcommand{\x}{\mathbf x}
\newcommand{\Y}{\mathbf Y}

\newcommand{\Z}{\mathbf Z}

\newcommand{\be}{\boldsymbol{\eta}}

\newcommand{\bt}{\boldsymbol{\theta}}

\DeclareMathOperator*{\argmax}{argmax}

\newcommand{\norm}[1]{\left\lVert#1\right\rVert}

\def\hat{\widehat}
\def\tilde{\widetilde}
\def\S{\mathbf S}

\newtheorem{proposition}{Proposition}

\def\spacingset#1{\renewcommand{\baselinestretch}%
{#1}\small\normalsize}

\title{Reducing Differential Item Functioning via Process Data}

\author{Ling Chen$^{*}$, Susu Zhang$^\dagger$, Jingchen Liu$^*$}

\date{$^*$Department of Statistics, Columbia University \& $^\dagger$Department of Psychology, University of Illinois Urbana-Champaign}

\begin{document}
\maketitle

\begin{abstract}
Testing fairness is a major concern in psychometric and educational research.
A typical approach for ensuring testing fairness is through differential item functioning (DIF) analysis.
DIF arises when a test item functions differently across subgroups that are typically defined by the respondents' demographic characteristics.
Most of the existing research has focused on the statistical detection of DIF, yet less attention has been given to reducing or eliminating DIF and understanding why it occurs. 
Simultaneously, the use of computer-based assessments has become increasingly popular. 
The data obtained from respondents interacting with an item are recorded in computer log files and are referred to as process data. 
Process data provide valuable insights into respondents’ problem-solving strategies and progress, offering new opportunities for DIF analysis. 
In this paper, we propose a novel method within the framework of generalized linear models (GLMs) that leverages process data to reduce and understand DIF. 
Specifically, we construct a nuisance trait surrogate with the features extracted from process data. 
With the constructed nuisance trait, we introduce a new scoring rule that incorporates respondents' behaviors captured through process data on top of the target latent trait.
We demonstrate the efficiency of our approach through extensive simulation experiments and an application to thirteen Problem Solving in Technology-Rich Environments (PSTRE) items from the 2012 Programme for the International Assessment of Adult Competencies (PIAAC) assessment.

\end{abstract}


\spacingset{1.7}
\section{Introduction}

Ensuring testing fairness in educational and psychometric assessments has been a major concern for researchers.
A typical approach to ensuring testing fairness is through differential item functioning \cite[DIF, ][]{holland2012differential} analysis. 
DIF occurs when an item's response function depends on not only the target latent trait to be measured by the item, but also the respondents' group memberships that are often linked to their demographic characteristics. 
When DIF is present, the measurement properties of the item differ systematically across groups, leading to measurement bias \citep{millsap2012statistical}.

The research on DIF has predominantly focused on the statistical detection of its existence. 
DIF detection methods are typically categorized as non-parametric \citep{holland1986differential, dorans86} and parametric \citep{rudner1980monte, raju1988area, lord1980applications, thissen2013use, lord1980applications, swaminathan1990detecting}.
More recently, DIF detection methods that do not require predefined group memberships or anchor items have been proposed \citep{chen2023dif, wallin2024dif, halpin2024differential}. 
While research has explored methods on handling items with DIF \citep{cho2016after, liu2022treatments}, items identified with significant DIF are often removed during item calibration, leading to wasted resources and efforts in their development and administration. 
Consequently, there has been growing interest in reducing or eliminating DIF, as well as
understanding the underlying reasons for why DIF occurs \citep{ackerman2024examining}. 

One way to attribute the cause of DIF is multidimensionality, where DIF arises due to the presence of secondary dimensions in the latent space \citep{kok1988item, ackerman1992didactic, shealy1993model}. 
Ideally, differences in the response probabilities solely reflect variations in the latent ability that the item is designed to assess, which is the primary dimension. 
However, secondary latent traits with heterogeneous distributions across subpopulations can also contribute to these differences. 
These secondary dimensions are called auxiliary if they are intentionally measured by the item, or nuisance otherwise \citep{roussos1996multidimensionality}.
The multidimensional IRT (MIRT) model has been used to analyze DIF, where both the target trait and the nuisance trait are used to model item response probabilities.
Latent DIF models have also been used to investigate the secondary dimensions, using mixture models to identify latent groups as the secondary dimension and associating it with examinees' demographic characteristics \citep{cohen2005mixture, de2011explanatory}.
The multiple-indicator multiple-cause (MIMIC) model provides another approach from the dimensionality perspective, although only the primary dimension is used \citep{de2011explanatory}.
In a mediated MIMIC model proposed in \cite{cheng2016mediated}, a secondary dimension construct (the scale of self-confidence) is used as a potential mediator.
Despite these advances, studies that rely only on response data face challenges, particularly when prior knowledge of nuisance traits is limited.
Thus, it is of interest to identify the secondary dimensions from data sources beyond the response outcome data.

One data source that presents new opportunities for identifying secondary latent dimensions in DIF analysis is process data \citep{he2021leveraging, wang2023subtask, zhang2023accurate, li2024exploring}. 
Process data capture the problem-solving processes as respondents interact with computer-based test items. The respondents' actions are logged as time-stamped action sequences in computer log files, making process data a detailed record of respondents' behaviors.
Two prominent examples of process data are from the Programme for International Student Assessment (PISA) and the Programme for the International Assessment of Adult Competencies (PIAAC). 
These assessments not only measure skills traditionally tested with paper-and-pencil methods but also evaluate more complex abilities such as problem-solving in technology-rich environments. 
Compared to traditional outcome data that are typically dichotomous (correct/incorrect) or polytomous (partial credit), process data provides more comprehensive insights into the strategies respondents use to solve problems, which highlights its potential in identifying nuisance traits. 
For example, engagement is an acknowledged nuisance trait, and the first principal component of features extracted from the 2012 PIAAC process data is highly correlated with engagement \citep{tang2020latent}.
If we are able to identify the secondary dimensions that lead to DIF via process data, we can reduce DIF by proposing a new scoring rule based on process data, and interpret the mechanism of DIF by identifying the heterogeneous behavioral patterns in process data that drive these differences.

In this paper, we propose a novel method for reducing DIF that leverages process data and introduce a corresponding scoring rule that only depends on the respondents' behaviors. We attribute DIF to multidimensionality and discuss the method within the framework of generalized linear models (GLM). 
We assume that the target latent trait (primary dimension) is known or has been estimated, while the nuisance trait is derived from process data.
The key innovation of our approach lies in constructing a surrogate for the nuisance trait using features extracted from process data. 
This surrogate is formulated as a linear combination of the process data features, with the weights determined by minimizing the maximum likelihood difference between models with and without the grouping variable. 
The motivation is to construct a measurement model with the target trait and nuisance trait surrogate, while minimizing the impact of the grouping variable on the measurement model.
We show that in the simple case of linear model (classical test theory), the proposed optimization problem has a closed-form solution that determines the optimal feature weights. 
In more complex scenarios involving multiple grouping variables or nonlinear models, the optimal solution can be found through numerical optimization techniques.
With the optimal feature weights, we propose a new scoring rule that incorporates both the target latent trait and nuisance trait surrogate that reduces or corrects DIF. 
We stress that the scoring rule is purely based on respondents' responses.
The effectiveness of this method is demonstrated through simulation studies and a case study. 
Additionally, we offer interpretations of the group differences based on their action sequences in the case study, providing an insight of understanding why DIF occurs.

The rest of the paper is structured as follows.
Section \ref{sec: method} outlines the methodology of the proposed approach, while Section \ref{sec: sim} presents the results of the simulation experiments. 
In Section \ref{sec: case}, we demonstrate a case study using the PIAAC 2012 dataset. Finally, we provide a discussion in Section \ref{sec: discussion}.

\section{Method}\label{sec: method}

Consider $N$ independent respondents and their process and outcome responses to one item of interest.
Let $Y_{i}\in\{0, \dots, C-1\}$ represent the response of the $i$-th respondent, where $C$ is the number of possible responses. For example, when $C=2$, $Y_{i}=1$ indicates correct response and $Y_{i}=0$ indicates otherwise. 
We use $\Y=(Y_{1},\dots, Y_{N})^\top\in\mathbb{R}^N$ to denote the vector of responses from all $N$ respondents. 
We introduce one grouping variable $Z_{i}\in\{0, 1\}$, where $Z_{i}=0$ represents the reference group and $Z_{i}=1$ represents the focal group.
The item is assumed to measure a unidimensional latent trait, denoted by $\theta_i$ for respondent $i\in[N]$. Let $\bo\theta=(\theta_1,\dots, \theta_N)^\top$ be the collection of latent traits of all respondents.
Additionally, we extract process features $\X = (\x_{1}, \dots, \x_{N})^\top \in\mathbb{R}^{N\times K}$ from the action sequences of the process data with the multidimensional scaling procedure proposed in \cite{tang2020latent}, where $\x_{i}\in\mathbb{R}^K$. 
Since process features capture most of the useful information of process data when the feature dimension is sufficiently large \citep{tang2020latent, tang2020auto}, we will use process features as a proxies for the original process data in this work.

\subsection{Differential Item Functioning}

We adopt a multidimensionality-based DIF framework. In addition to the target trait dimension, $\theta_i$, a nuisance trait dimension, $\eta_{i}$, also influences the probability of the response. 
Denote $\bo\eta=(\eta_{1},\dots, \eta_{N})^\top$ as the vector encompassing the nuisance traits for all respondents.

DIF occurs when the item response probability depends on the nuisance trait, and there is distributional difference of the nuisance trait among the two groups. Conditional on $\theta_i$, $\eta_{i}$ and $Z_i$, it is assumed that $Y_{i}$ are independently distributed. 
When the distribution of $\eta_{i}$ conditional on $\theta_i$ differs across the two subgroups, we have
\begin{align*}
    p(Y_{i}=y|\theta_i, Z_i=0) = & \int p(Y_{i}=y|\theta_i, \eta_{i}) \cdot p(\eta_{i}|\theta_i, Z_i=0) d\eta_{i}  \\
    \neq & \int p(Y_{i}=y|\theta_i, \eta_{i}) \cdot p(\eta_{i}|\theta_i, Z_i=1) d\eta_{i} = p(Y_{i}=y|\theta_i, Z_i=1).
\end{align*}
Therefore, under this multidimensional framework, a uni-dimensional measurement model leads to DIF. 

Specifically, we consider a GLM with the following conditional distribution:
\begin{equation}\label{eq:glm}
    Y_{i} \sim p(y | \mu_{i}), \quad \text{where}\ \  g(\mu_{i}) = d + a_{0} \theta_i + a_{1} \eta_{i} + \lambda Z_i.
\end{equation}
Here $g(\cdot)$ is the link function with respect to the response mean $\mu_{i}=\mathbb{E}[Y_{i} | \theta_i, \eta_{i}, Z_i]$, and $d, \ba:= (a_{0}, a_{1}), \lambda$ are unknown coefficients. When $\lambda=0$, the distributional difference of $\eta_{i}|\theta_i$ among the two subgroups is the only DIF source in a uni-dimensional measurement model.
The model specified by \eqref{eq:glm} is quite general, as we allow $g(\cdot)$ to take a general form for a wide range of response types such as binary, polytomous, and continuous responses.
For binary responses, the logistic regression model with $g(\mu_{i})=\ln\left(\mu_{i} / (1-\mu_{i})\right)$ is referred to as the multidimensional two-parameter logistic (M2PL) IRT model. 
In addition, $g(\mu_{i})=\Phi^{-1}(\mu_{i})$ corresponds to the probit regression model with $\Phi(\cdot)$ being the cumulative distribution function of the standard normal distribution.
When $g(\mu_i)=\mu_i$, the model becomes a linear regression model for continuous responses. 
Also note that the model setup can be easily extended to multiple target traits and multiple nuisance traits. For the ease of exposition, we focus on the case with only one target trait and one nuisance trait.

\subsection{DIF Reduction}
One of the primary challenges in applying the multidimensionality-based DIF analysis is that the nuisance trait, $\eta_{i}$, is unobserved and cannot be directly measured.
We propose to construct a surrogate for the unobserved nuisance trait in a way that corrects DIF within the GLM framework.
Specifically, we aim to build a surrogate, $\hat\eta_{i}$, such that DIF is only attributed to the distributional differences in $\hat\eta_{i}|\theta_i$, and the conditional probabilities of item responses become approximately equal across different subgroups. Formally, we aim to satisfy the following condition:
\begin{equation}\label{eq:no_dif}
    p(Y_{i}=y | \theta_i, \hat\eta_{i}, Z_i=0)\approx p(Y_{i}=y| \theta_i, \hat\eta_{i}, Z_i=1).
\end{equation}

To achieve \eqref{eq:no_dif}, we propose leveraging process data to construct the nuisance trait surrogate, driven by two key motivations. 
Firstly, process data capture the entire sequence of actions taken by each respondent as they interact with and solve an item, providing a rich source of information on various nuisance traits. 
For instance, the first principal component of process features derived from process data has been shown to correlate strongly with engagement, a known nuisance trait that influences item responses \citep{tang2020latent}.
Second, process data typically predict the final response with perfect accuracy. 
By analyzing the respondent’s full sequence of actions, we can infer whether they answered the item correctly or their partial scores. 
In theory, adding all available process features in the model eliminates DIF. 
Yet this approach is not ideal as the resulting measurement model solely depends on the process features and provides little, if any, information of the target trait.

The core of our proposed method is to identify an optimal linear combination of process features as a surrogate for the nuisance trait $\eta_{i}$. 
Specifically, we consider
$$
    \eta_{i}=\bw^\top \x_{i},
$$
where $\bw=(\omega_{1}, \dots, \omega_{K})^\top \in\mathbb{R}^K$ is the weight vector and we assume $\|\bw\|=1$ for model identifiability.
Our objective in identifying $\bw$ is to minimize a quantity equivalent to the likelihood ratio test statistic,
\begin{equation}\label{eq:objective}
    L(\bw) := \max_{d, \ba, \lambda} l(d, \ba, \lambda) - \max_{d, \ba, \lambda=0} l(d, \ba, \lambda),
\end{equation}
where $l(\cdot)$ is the log-likelihood function,
\begin{equation}\label{eq:likelihood}
    l(d, \ba, \lambda) = \sum_{i=1}^N \log p(Y_{i}|\theta_i, \eta_{i}, Z_i).
\end{equation}

Function \eqref{eq:objective} quantifies how much model fit is increased after adding the grouping variable into the model, thus can be viewed as a quantification of the DIF effect. 
When there is no DIF exhibited, adding the grouping variable into the model would barely increase the likelihood, and we would expect $L(\bw)$ to be close to $0$. 
This objective function enables us to optimize $\bw$ by comparing models with and without the grouping variable, ultimately reducing or removing DIF.
Therefore, we propose the estimation of $\bw$ to be the minimizer of the objective function
\begin{equation}\label{eq:minimizer}
    \hat \bw = \arg\min_{\norm{\bw}=1} L(\bw).
\end{equation}
and the nuisance trait surrogate is $\hat\eta_{i}=\hat\bw^\top\x_{i}$.
We will show that the solution of \eqref{eq:minimizer} has a closed-form solution in the linear regression model with one grouping variable; see Section \ref{sec: linear}. 
In other cases, we can optimize the objective function numerically. 

\begin{figure}[t!]
    \centering
    \includegraphics[width=0.7\linewidth]{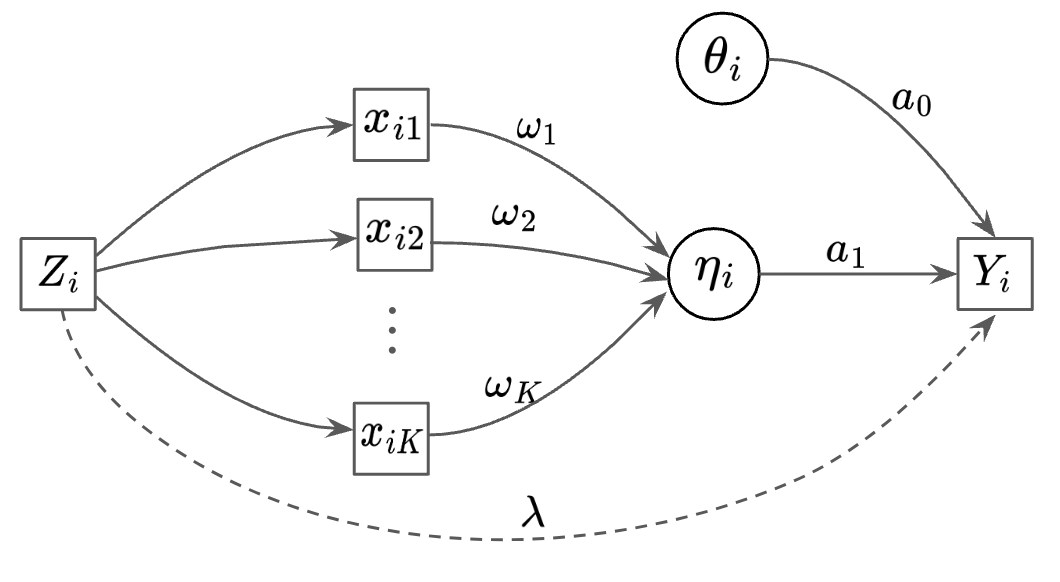}
    \caption{Measurement model without the intercept. 
    }
    \label{fig:structure}
\end{figure}

Figure \ref{fig:structure} displays the updated measurement model incorporating both the nuisance trait and target trait. 
With the nuisance trait surrogate, we update the initial estimate of the target trait with the two-dimensional measurement model. 
Consider the case with $J$ items, among which items $j\in\mathcal{B}\subset[J]$ exhibit DIF.
Suppose the nuisance trait surrogates $\hat\be_j=(\hat\eta_{ij})\in\mathbb{R}^N, j\in\mathcal{B}$ have been estimated for the DIF items, then we obtain the MLE of the item parameters $\hat{d}_j, \hat{a}_{0j}, \hat{a}_{1j}$ with $\lambda_j$ fixed as $0$ in \eqref{eq:glm} for $j\in\mathcal{B}$, and $\hat{d}_j, \hat{a}_{0j}$ with $a_{1j}, \lambda_j$ fixed as $0$ for $j\in[J]\backslash \mathcal{B}$. 
The target trait estimate is then updated by the maximum likelihood estimate (MLE)
\begin{equation}\label{eq:update}
    \hat\theta_i = \argmax_{\theta} \sum_{j=1}^J \log p_{ij}(\theta),
\end{equation}
where 
\begin{align*}
    p_{ij}(\theta) & = p(Y_{ij}|\theta, \hat\eta_{ij}, \hat d_j, \hat{a}_{0j}, \hat{a}_{1j}, \lambda_j=0), \quad j\in\mathcal{B}, \\
    p_{ij}(\theta) & = p(Y_{ij}|\theta, \hat d_j, \hat{a}_{0j}, a_{1j}=0, \lambda_j=0), \quad j\in [J]\backslash\mathcal{B}.
\end{align*}

\subsection{A Special Case: Linear Model with Closed-form Solution} \label{sec: linear}
When the link function $g$ is the identity function, model \eqref{eq:glm} becomes the linear regression model.
As DIF is defined as the group difference of the distribution of $Y$ conditional on the latent trait, we conduct our DIF analysis in the orthogonal subspace of $\bo\theta$ in the linear model. 
To be more specific, we consider the residuals of $\Y, \Z, \X$ after regressing on $(1, \bo\theta)$ respectively, denoted by $\Y^\dagger, \Z^\dagger, \X^\dagger$. 

The model with and without the grouping variable can be rewritten as a reduced and a full linear regression model
\begin{align}
    \Y^\dagger & = a_{1}' \bo\eta + \bo\epsilon, \label{eq:7}\\
    \Y^\dagger & = a_{1} \bo\eta + \lambda \Z^\dagger + \bo\delta, \label{eq:8}
\end{align}
where $\bo\eta = \X^\dagger \bw$. If we assume $\epsilon_{i}\stackrel{i.i.d}{\sim} N(0, \sigma_{\epsilon}^2)$ and $\delta_{i}\stackrel{i.i.d}{\sim} N(0, \sigma_{\delta}^2)$, the objective function \eqref{eq:objective} is equivalent to
\begin{equation}\label{eq:objective_linear}
    L(\bw) = \frac{n}{2} \log(\|\hat{\bo\epsilon}\|^2) - \frac{n}{2} \log(\|\hat{\bo\delta}\|^2),
\end{equation}
where $\hat{\bo\epsilon}$ and $\hat{\bo\delta}$ are the linear regression residuals in \eqref{eq:7} and \eqref{eq:8}. 
The zero of the objective \eqref{eq:objective_linear} turns out to have a closed form expression under some weak conditions, in which case DIF can be fully removed.
Without loss of generality, we assume that all the features are orthogonal and scaled, i.e. $\X^{\dagger\top}\X^\dagger=\mathbf{I}_K$. 
This is achieved by principal component analysis in practice. 
We also assume that $\Y^{\dagger\top}\Z^\dagger>0$. 
\begin{proposition}\label{prop}
    Assume that $\X^{\dagger\top}\X^\dagger=\mathbf{I}_K$. Let $\hat\Y = \X^{\dagger\top} \Y^\dagger$ and $\hat\Z = \X^{\dagger\top} \Z^\dagger$.
    Under the condition that 
    \begin{equation}\label{eq:condition}
        \frac{-\|\hat\Y\|\|\hat\Z\|+\hat\Y^\top \hat\Z}{2} < \Y^{\dagger\top}\Z^\dagger < \frac{\|\hat\Y\|\|\hat\Z\| + \hat\Y^\top \hat\Z}{2},
    \end{equation}
   there exists $\hat\bw$ such that $\norm{\hat\bw}=1$ and $L(\hat\bw)=0$.
   Specifically, denote
\begin{align}
    \alpha = \sqrt{\frac{2\Y^{\dagger\top}\Z^\dagger + \|\hat\Y\|\|\hat\Z\| - \hat\Y^\top \hat\Z}{2\|\hat\Y\|\|\hat\Z\|}} & , \ \beta = \sqrt{\frac{-2\Y^{\dagger\top}\Z^\dagger + \|\hat\Y\|\|\hat\Z\| + \hat\Y^\top \hat\Z}{2\|\hat\Y\|\|\hat\Z\|}}, \notag \\
    \bo q_1 = \frac{\|\hat\Y\|\hat\Z + \|\hat\Z\|\hat\Y}{\left\|\|\hat\Y\|\hat\Z + \|\hat\Z\|\hat\Y \right\|} & , \  \bo q_2= \frac{\|\hat\Y\|\hat\Z - \|\hat\Z\|\hat\Y}{\left\| \|\hat\Y\|\hat\Z - \|\hat\Z\|\hat\Y \right\|}. \notag
\end{align}
Then $\hat\bw = \alpha \bo q_1 + \beta \bo q_2$ or $\hat\bw = \alpha \bo q_1 - \beta \bo q_2$ satisfies $L(\hat\bw)=0$.
\end{proposition}
We note that $L(\bw)$ has multiple zeros, as established in Proposition \ref{prop}. We choose $\hat\bw_1=\alpha \bo q_1 + \beta \bo q_2$ over $\hat\bw_2=\alpha \bo q_1 - \beta \bo q_2$ for the following reason. When $\X^\dagger$ predicts $\Y^\dagger$ with high accuracy, $\X^\dagger \hat\bw_1$ aligns with the projection of $\Z^\dagger$ onto the column space of $\X^\dagger$, whereas $\X^\dagger \hat\bw_2$ aligns with the projection of $\Y^\dagger$ onto the same subspace. As the goal is to reduce DIF, we choose to use $\hat\bw_1$ over $\hat\bw_2$. For further details, see the proof of Proposition \ref{prop} in the Appendix.

There are several scenarios in which condition \eqref{eq:condition} holds.
The first scenario occurs when $\Y^{\dagger\top} \Z^\dagger= 0$, which arises if the response and the grouping variable are independent conditional on the target trait, indicating that the item is not a DIF item, and condition \eqref{eq:condition} is automatically satisfied. 
The second scenario is when $\X^\dagger$ has high linear predictability of $\Y^\dagger$.
Specifically, when $\X^\dagger$ predicts $\Y^\dagger$ with high accuracy, we find that $\X^\dagger\X^{\dagger\top}\Y^\dagger \approx \Y^\dagger$, which leads to $\hat\Y^\top\hat\Z=\Y^{\dagger\top}\X^\dagger\X^{\dagger\top}\Z^\dagger \approx \Y^{\dagger\top} \Z^\dagger$, making it straightforward to verify condition \eqref{eq:condition}. 
This is the case where process data capture the nuisance factors that affect the response. In the simplest case, we assume $\X$ can linearly predict $\Y$ with full accuracy and let $\Y = \X \A$ with some vector $\A\in\mathbb{R}^K$. Then, $$\Y^\dagger = \Y-\mathbb{E}[\Y|\bo\theta] = \left( \X-\mathbb{E}[\X|\bo\theta]\right)\A =  \X^\dagger \A.$$
Accordingly, we consider this case to be achievable.
The third scenario is when $\X$ has high linear predictability of $\Z$. Following similar calculations, we conclude that condition \eqref{eq:condition} is satisfied. With that being said, it is generally difficult to predict $\Z$ with process data.

\subsection{General Cases}
While we have previously focused on uniform DIF in a linear factor model with one grouping variable, the proposed method is applicable to other general cases. 
In what follows, we will shift our focus to addressing non-uniform DIF, continuous covariates, multiple grouping variables
, and nonlinear models. The goal is to minimize the objective function \eqref{eq:objective}.

\paragraph{Non-uniform DIF.}
Non-uniform DIF occurs when not only the intercept parameter, but also the discrimination parameter (the coefficient of $\theta_i$) differs across groups. 
More specifically, for non-uniform DIF, \eqref{eq:glm} becomes
\begin{equation}\label{eq:linear-nonuniform}
    g(\mu_{i}) = d + a_{0} \theta_i + a_{1} \eta_{i} + \lambda Z_i + \lambda' Z_i\theta_i.
\end{equation}
Non-uniform DIF can be viewed as a special case involving one grouping variable $Z_i$ and a continuous covariate $Z_i\theta_i$. Therefore, non-uniform DIF is included in the continuous covariates and multiple-group cases discussed below.

\paragraph{Continuous covariates.}
In some applications, DIF is brought by continuous covariates. For instance, in computer-based tests, age is a very important variable.
As our proposed method does not require $Z_i$ to be a discrete variable, it is applicable when $Z_i$ is a continuous variable. To see this, let $Z_i\in\mathbb{R}$, and we aim to construct $\hat{\eta}_{ij}$ such that
\begin{equation}\label{eq:no_dif_continuous}
    p(Y_{i}=y | \theta_i, \hat\eta_{i}, Z_i)\approx p(Y_{i}=y| \theta_i, \hat\eta_{i}).
\end{equation}
When \eqref{eq:no_dif_continuous} holds, we expect the objective function \eqref{eq:objective} to be close to $0$. Therefore, minimizing \eqref{eq:objective} to reduce DIF is valid for continuous covariates.

\paragraph{Nonlinear models.}
When the link function $g(\cdot)$ is not linear, e.g. the M2PL model and the Probit regression model, minimizing the objective function \eqref{eq:objective} is a nested optimization problem that takes in different forms depending on the model employed. Again, we rely on numerical methods to approximate the solution.

\paragraph{Multiple grouping variables.}
Sometimes it is of interest to evaluating DIF over more than one grouping variables \citep{kim1995detection, bauer2020simplifying}.
For cases involving multiple grouping variables, the expression of the objective function in terms of $\bw$ becomes significantly more complex. To address this, we propose an approximation of the objective function for $M$ grouping variables $\Z_1,\dots, \Z_M\in\mathbb{R}^N, M\ge 2$ 
as follows:
\begin{equation}\label{eq:objective_multiple}
    L(\bw) = \sum_{m=1}^M L^{(m)}(\bw),
\end{equation}
where $L^{(m)}(\bw)$ is the objective function \eqref{eq:objective} corresponding to $\Z_m$. 
In general, there is no closed-form solution for the minimizer of \eqref{eq:objective} or \eqref{eq:objective_multiple} with the presence of multiple grouping variables. Therefore, we rely on numerical methods to approximate the minimizer.

\subsection{Procedure}
We outline the procedure of the proposed method in a practical setting with $J$ items. 
\vspace{-0.5em}
\begin{enumerate}
    \item Suppose we have access to a set of anchor items that are DIF-free, which are used to perform DIF detection on all the items. Suppose DIF has been detected on a subset of items $\mathcal{B}\subset [J]$.
    \item Obtain an initial latent trait estimate $\hat\bt^{(0)}$ using the items without DIF.
    \item For each item, perform the proposed method to obtain the nuisance trait surrogates and DIF-corrected model parameters. More specifically, for each item $j\in\mathcal{B}$,
    \begin{enumerate}
        \item With the initial ability estimate, $\hat\bt^{(0)}$, find the minimizer $\hat\bw_j$ in Equation \eqref{eq:minimizer} and obtain the nuisance trait estimate $\hat{\bo\eta}_j=\X_j\hat\bw_j$.
        \item With $\hat\bt^{(0)}, \hat{\bo\eta}_j$, obtain the item parameter estimates $\hat d_j, \hat{a}_{0j}, \hat{a}_{1j}$ in \eqref{eq:glm} with $\lambda_j$ set to $0$, using the full data.
    \end{enumerate}
    \item Obtain the updated estimate of $\theta_i$ with \eqref{eq:update}, utilizing the nuisance trait surrogates and the calibrated measurement models.
\end{enumerate}


\section{Simulation Studies}\label{sec: sim}

We carry out extensive simulation experiments to evaluate the proposed method in this section. 
The goal is to show that the proposed method is able to minimize the objective function, accurately estimate the nuisance traits and the item parameters, and correct for target trait estimation from the DIF items. 

\subsection{Simulation Settings}
We consider three settings with the sample sizes $N=200, 500, 1000$.
Among the subjects, $2/3$ are in the reference group and $1/3$ are in the focal group.
We fix the number of items as $J=25$ and consider low, medium, high proportions of DIF items, that is, $5, 10, 15$ DIF items. 
We also consider two settings for the DIF parameters $a_{1j}$. For small DIF effects, $a_{1j}$s are uniformly sampled from $0.5$ to $1$; for large DIF effects, the range is from $1$ to $1.5$. 
In summary, there are $18$ simulation settings varying in sample size, proportion of DIF items, and DIF effect size.
For each simulation setting, $100$ independent replications are generated.
In each replication, we sample the target latent trait $\theta_i$ independently from the standard normal distribution. 
The difficulty parameters $d_j$ are sampled uniformly from $-1$ to $1$ and the discrimination parameters $a_{0j}$ are sampled uniformly from $1$ to $2$. 
As the generation of process data is challenging, we generate the process data features directly.
The number of process data features for each item is fixed at $K=10$.
The process data features $\x_{ij}\in\mathbb{R}^K$ are first independently sampled from the multivariate Gaussian distribution $N(\bo{\mu}^{(Z_i)}, \mathbf{I}_K)$. The mean of the process features depends on the respondent's group membership. For the reference group, $\bo{\mu}^{(0)}=\mathbf{1}_K$, and for the focal group $\bo{\mu}^{(1)}=-\mathbf{1}_K$. We then right multiply $\X_j$ with sample $\text{Cov}(\X_j)^{-1/2}$ for the process features to have the identity matrix as the covariance.
The ground truth nuisance trait is a linear combination of the process features: $\eta_{ij}=\bw_j^\top\x_{ij}$, with $\omega_{jk}$ sampled independently from the exponential distribution with rate $1$ and $\bw_j$'s are scaled to have unit norm. Note that the generated nuisance traits for each item have unit norm.
We consider both the linear model and the M2PL model in generating the item responses.
As DIF is solely introduced by the distributional difference of the nuisance trait in simulation, $\lambda_j$ is set to be $0$ in \eqref{eq:glm} when generating the item responses for both models. 


For the linear factor model, we set the variance of the item response noise as $1$ when generating the item responses. To make sure that the features can almost perfectly predict each item response in the linear model, we add one column of $\Y_j + N(0, 0.1\cdot \mathbf{I}_n)$ to the generated $\X_j$ for each item $j$ and generate the nuisance traits with the same manner as mentioned above. 
The initial target trait estimates $\hat\theta_i^{(0)}$ are obtained with factor analysis using responses from the DIF-free items.
To construct the nuisance trait, we adopt the closed form expression of $\hat\bw$ from Proposition \ref{prop}.
For the M2PL model, we generate $Y_{ij}$ with the link function $g(\cdot)$ being the the logit function. 
The initial target trait estimates $\hat\theta_i^{(0)}$ are the MLE estimates using the DIF-free items.
The nuisance traits are estimated by solving \eqref{eq:minimizer} using the \texttt{optim} function with the L-BFGS-B  optimizer in \texttt{R}.

The simulation above corresponds to uniform DIF. In addition, we consider the case with non-uniform DIF so that $\lambda'$ in \eqref{eq:linear-nonuniform} is not zero. Similar simulation settings are adopted except for the generation of process data features. To ensure the existence of non-uniform DIF, the process features $\x_{ij}$ are sampled from the multivariate Gaussian distribution $N(\bo{\mu}^{(Z_i)} + \gamma_j \theta_iZ_i, \mathbf{I}_K)$ with $\gamma_j$ simulated from the exponential distribution with rate $1$.

\subsection{Evaluation Criteria}
We consider five evaluation criteria. Firstly, we check wether the proposed method is able to reduce the objective function value. Specifically, we verify whether zero of \eqref{eq:objective} is obtained for the linear model with Proposition \ref{prop}. 
Secondly, we evaluate the correlation of nuisance trait estimation compared to its ground truth. 
Thirdly, we calculate the mean squared error (MSE) of the item parameter estimations. 
In addition, to evaluate measurement reliability after changing the scoring rule, we calculate the Fisher information (FI) of the target trait for the DIF items. 
For the linear model, target trait FI for the item $j\in\mathcal{B}$ is $\text{FI}_j = \hat{a}_{0j}^2 / \hat{\sigma}_j^2$,
where $\hat{a}_{0j}$ is the estimated coefficient for the target trait and $\hat{\sigma}_j^2$ is the estimated variance. For the M2PL model, $\text{FI}_j = \frac{1}{N} \sum_{i=1}^N \hat{p}_{ij}(1-\hat{p}_{ij}) \hat{a}_{0j}^2$, where $\hat{p}_{ij}$ is the estimated response of the logistic model. 
Last but not least, we consider the between-group sum of squared (SS) bias for target trait estimation using the DIF items. We elaborate more on this evaluation criterion assuming one grouping variable with a focal group and a reference group. 
Note that this criterion can be easily generalized to multiple grouping variables. For any target trait estimate $\tilde\bt$, define the bias $\bo\nu:=\tilde\bt-\bt$ and denote the mean of bias as $\bar\nu:=\frac{1}{N}\sum_{i=1}^N\nu_i$. Consider the between-group SS for the bias of $\tilde\bt$ in analysis of variance (ANOVA):
\begin{align}
    SSB_{\tilde\bt} = N_r \left(\frac{1}{N_r}\sum_{i\in\mathcal{N}_r}\nu_i - \bar\nu \right)^2 + N_f \left(\frac{1}{N_f}\sum_{i\in\mathcal{N}_f}\nu_i - \bar\nu \right)^2,
\end{align}
where $\mathcal{N}_r$ and $\mathcal{N}_f$ are the index sets of the reference group and focal group respectively, and $N_r=|\mathcal{N}_r|, N_f=|\mathcal{N}_f|$. To illustrate that the proposed method is able to de-bias target trait estimation, we compare the above-defined value for two estimates. The benchmark estimate is the MLE computed using the responses from the DIF items in $\mathcal{B}$, assuming DIF is not present:
\begin{align*}
    \check\theta_i = \argmax_{\theta} \sum_{j\in\mathcal{B}} \log p(Y_{ij} | \theta, \check d_j, \check{a}_{0j}, a_{1j}=0, \lambda_j=0),
\end{align*}
where $\check{d}_j, \check{a}_{0j}$ are calibrated on the DIF items. 
The second estimate is the MLE computed after DIF-correction:
\begin{align*}
    \hat\theta_i = \argmax_{\theta} \sum_{j\in\mathcal{B}} \log p(Y_{ij} | \theta, \hat d_j, \hat{a}_{0j}, \hat{a}_{1j}, \lambda_j=0),
\end{align*}
where $\hat{d}_j, \hat{a}_{0j}, \hat{a}_{1j}$ are calibrated on the DIF items with the nuisance trait surrogates.
Because of the presence of DIF, $\check\bt$ is expected to be over-estimated within one group, and under-estimated within the other, leading to large values of $SSB_{\check\bt}$. With the proposed method, we expect $SSB_{\hat\bt}$ for the corrected estimate to be small compared to $SSB_{\check\bt}$ for the not-corrected estimate.

\subsection{Simulation Results}

\begin{figure}[b!]
    \centering
    \includegraphics[width=0.8\linewidth]{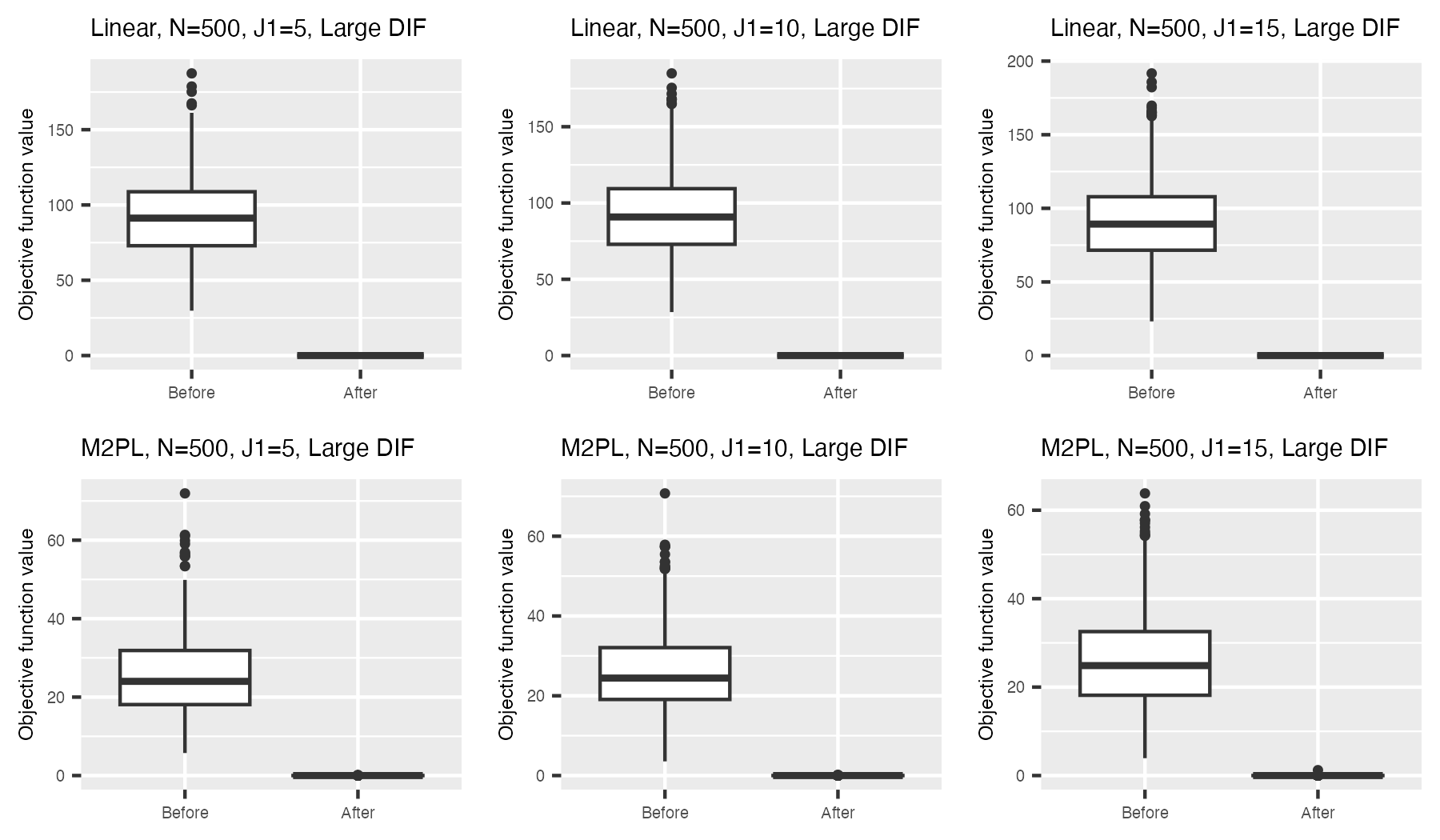}
    \caption{Values of the objective function before and after adding the nuisance trait surrogate for the linear model (upper) and the M2PL model (lower) with \emph{uniform} DIF, $N=500$, and large DIF.}
    \label{fig:sim_obj}
\end{figure}

\begin{figure}[hbt!]
    \centering
    \includegraphics[width=0.8\linewidth]{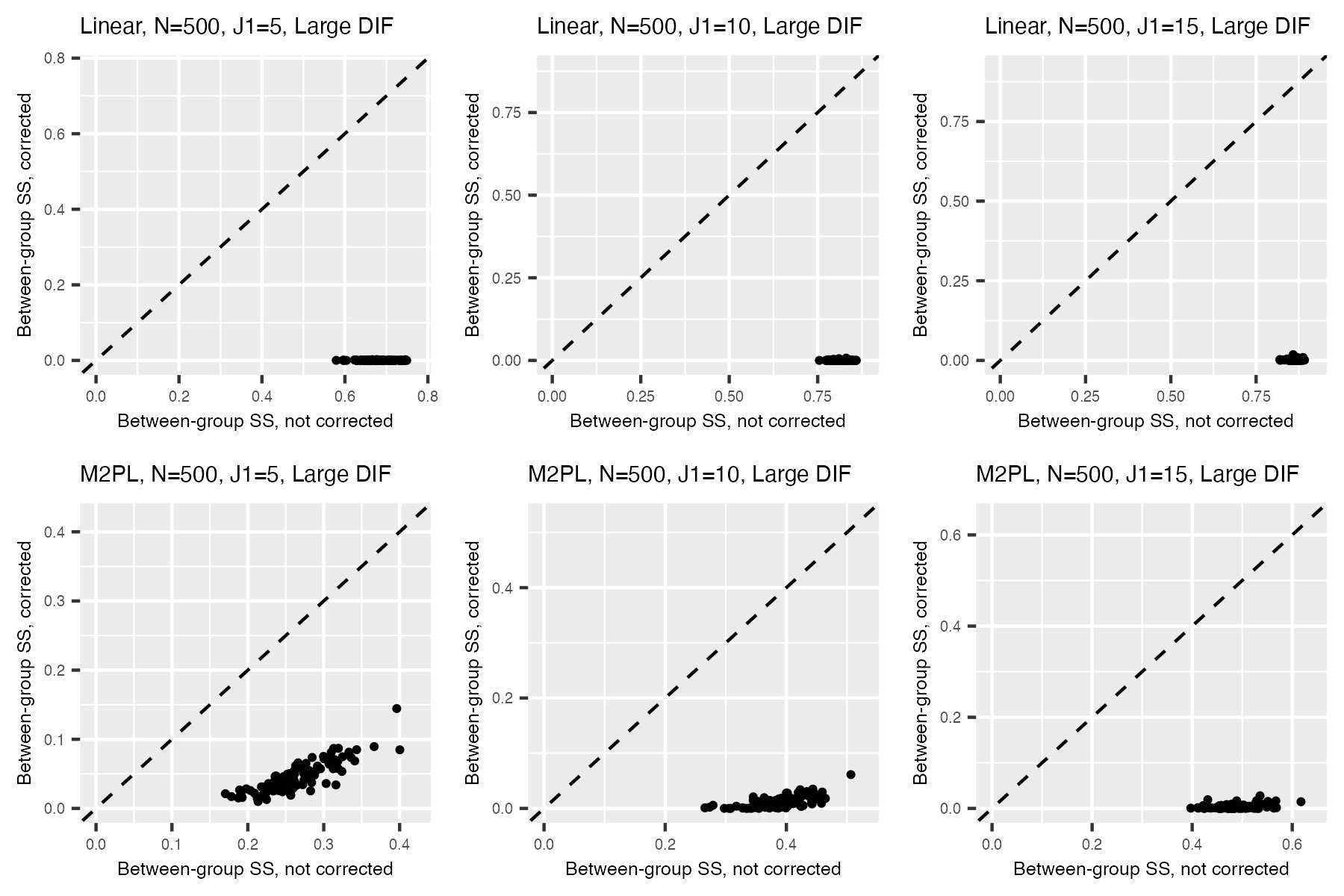}
    \caption{Between-group sum of squared bias for target trait estimation for the linear model (upper) and the M2PL model (lower) with \emph{uniform} DIF, $N=500$, and large DIF. The x-axis corresponds to the estimation without DIF correction using the DIF items; the y-axis corresponds to the DIF-corrected estimation using the DIF items.}
    \label{fig:anova_uniform_N500}
\end{figure}

Figure \ref{fig:sim_obj} summarizes the values for the objective function \ref{eq:objective} before and after adding the nuisance trait surrogate for the linear and M2PL models, with uniform DIF, $N=500$, and large DIF.
For complete results of all simulation settings, see Figures \ref{fig:sim_obj_linear} and \ref{fig:sim_obj_m2pl} in the Appendix.
We are able to minimize the objective function across simulation settings and replications. Specifically, we are able to obtain the zero of the objective function for the linear model. 
Tables \ref{tab:mse_linear} and \ref{tab:mse_m2pl} demonstrate the MSE of the item parameter estimates and the correlation between the ground truth and estimated nuisance traits. 
We observe that the MSE values are small in magnitude and the nuisance trait estimation correlation is high. 
It also shows that as the sample size increases, MSE tends to decrease and the nuisance trait correlation tends to increase. 
On the other hand, larger DIF effects and DIF item proportion lead to larger MSE. 
For the linear model, the nuisance trait correlation does not change with the sample size. For the M2PL model, it increases with the sample size.
Figures \ref{fig:sim_info_linear_uniform} and \ref{fig:sim_info_M2PL_uniform} in the Appendix summarize the Fisher information of the target trait $\theta$ before and after adding the nuisance trait surrogate for the linear and M2PL models respectively. We observe an increase in Fisher information of the target trait in the measurement model after correcting for DIF in both models.
Furthermore, Figure \ref{fig:anova_uniform_N500} compares the between-group SS bias of the corrected and not-corrected target trait estimates using the DIF items, for the linear and M2PL models. The x-axis corresponds to the estimation without DIF correction, and the y-axis corresponds to the DIF-corrected estimation. The corresponding simulation setting is $N=500$ with large DIF; for the complete results, see Figures \ref{fig:sim_anova_linear_uniform} and \ref{fig:sim_anova_M2PL_uniform} in the Appendix. We see that the proposed method is able to correct for the target estimation bias introduced by DIF, as the between-group SS bias is significantly reduced after DIF correction for both models.

\begin{table}[hbt!]
\footnotesize
\centering
\begin{tabular}{cccccccc}
\hline
& & \multicolumn{3}{c}{Small DIF} & \multicolumn{3}{c}{Large DIF} \\ \cmidrule(lr){3-5} \cmidrule(lr){6-8}
\multicolumn{2}{c}{} & $J_1$=5 & $J_1$=10 & $J_1$=15 & $J_1$=5 & $J_1$=10 & $J_1$=15 \\ 
\hline
\multirow{3}{*}{MSE ($d$)} & $N=200$ & 0.0618 & 0.0507 & 0.0665 & 0.1206 & 0.1249 & 0.1276 \\
 & $N=500$ & 0.0401 & 0.0451 & 0.0512 & 0.1085 & 0.1133 & 0.1122 \\
 & $N=1000$ & 0.0423 & 0.0370 & 0.0387 & 0.1042 & 0.1122 & 0.1083 \\ \hline
\multirow{3}{*}{MSE ($a_0$)} & $N=200$ & 0.0140 & 0.0155 & 0.0144 & 0.0209 & 0.0177 & 0.0225 \\
 & $N=500$ & 0.0059 & 0.0057 & 0.0053 & 0.0078 & 0.0071 & 0.0079 \\
 & $N=1000$ & 0.0026 & 0.0030 & 0.0026 & 0.0029 & 0.0036 & 0.0035 \\ \hline
\multirow{3}{*}{MSE ($a_1$)} & $N=200$ & 0.0465 & 0.0448 & 0.0460 & 0.1099 & 0.1111 & 0.1114 \\
 & $N=500$ & 0.0389 & 0.0401 & 0.0394 & 0.1052 & 0.1030 & 0.1043 \\
 & $N=1000$ & 0.0409 & 0.0392 & 0.0392 & 0.1053 & 0.1012 & 0.1051 \\ \hline
 \multirow{3}{*}{Corr ($\hat\be, \be$)} & $N=200$ & 0.7540 & 0.7542 & 0.7544 & 0.7609 & 0.7582 & 0.7573 \\
& $N=500$ & 0.7598 & 0.7575 & 0.7574 & 0.7577 & 0.7609 & 0.7585 \\
 & $N=1000$ & 0.7549 & 0.7568 & 0.7586 & 0.7566 & 0.7616 & 0.7560 \\ \hline
\end{tabular}
\caption{Mean squared error of item parameter estimates and nuisance trait correlation for the \emph{linear model} with \emph{uniform} DIF under different simulation settings. The values are averaged across the DIF items and replications.}
\label{tab:mse_linear}
\end{table}

\begin{table}[hbt!]
\footnotesize
\centering
\begin{tabular}{cccccccc}
\hline
 & & \multicolumn{3}{c}{Small DIF} & \multicolumn{3}{c}{Large DIF} \\ \cmidrule(lr){3-5} \cmidrule(lr){6-8}
\multicolumn{2}{c}{} & $J_1$=5 & $J_1$=10 & $J_1$=15 & $J_1$=5 & $J_1$=10 & $J_1$=15 \\ 
\hline
\multirow{3}{*}{MSE ($d$)} & $N=200$ & 0.0447 & 0.0530 & 0.0444 & 0.0505 & 0.0574 & 0.0552 \\
 & $N=500$ & 0.0191 & 0.0176 & 0.0190 & 0.0222 & 0.0231 & 0.0222 \\
 & $N=1000$ & 0.0108 & 0.0101 & 0.0112 & 0.0120 & 0.0116 & 0.0132 \\ \hline
\multirow{3}{*}{MSE ($a_0$)} & $N=200$ & 0.0794 & 0.0799 & 0.0821 & 0.0851 & 0.0894 & 0.0971 \\
 & $N=500$ & 0.0367 & 0.0435 & 0.0496 & 0.0454 & 0.0541 & 0.0715 \\
 & $N=1000$ & 0.0275 & 0.0330 & 0.0474 & 0.0393 & 0.0447 & 0.0577 \\ \hline
\multirow{3}{*}{MSE ($a_1$)}  & $N=200$ & 0.0603 & 0.0526 & 0.0598 & 0.0979 & 0.1040 & 0.1111 \\
 & $N=500$ & 0.0258 & 0.0302 & 0.0310 & 0.0775 & 0.0812 & 0.0908 \\
 & $N=1000$ & 0.0272 & 0.0263 & 0.0266 & 0.0748 & 0.0744 & 0.0884 \\ \hline
 \multirow{3}{*}{Corr ($\hat\be, \be$)} & $N=200$ & 0.7478 & 0.7489 & 0.7432 & 0.8264 & 0.8219 & 0.8242 \\
 & $N=500$ & 0.8322 & 0.8215 & 0.8231 & 0.8654 & 0.8669 & 0.8631 \\
 & $N=1000$ & 0.8487 & 0.8511 & 0.8568 & 0.8795 & 0.8819 & 0.8773 \\ \hline
\end{tabular}
\caption{Mean squared error of item parameter estimates and nuisance trait correlation for the \emph{M2PL model} with \emph{uniform} DIF under different simulation settings. The values are the averaged across the DIF items and replications.}
\label{tab:mse_m2pl}
\end{table}

Results for non-uniform DIF are deferred to the Appendix. In Figures \ref{fig:sim_obj_linear_nonuniform}, \ref{fig:sim_obj_m2pl_nonuniform}, we observe an increase in the minimized objective function values for non-uniform compared to uniform DIF, although the proposed method is still able to reduce DIF significantly. 
Similar results for item parameter and nuisance trait estimation, target trait Fisher information, and between-group SS bias are observed; see Tables \ref{tab:mse_linear_nonuniform}, \ref{tab:mse_m2pl_nonuniform}, Figures \ref{fig:sim_info_linear_nonuniform}, \ref{fig:sim_info_m2pl_nonuniform}, \ref{fig:sim_anova_linear_nonuniform}, \ref{fig:sim_anova_m2pl_nonuniform} in the Appendix.


\section{Case Study}\label{sec: case}
We use the PIAAC 2012 survey data as a case study to demonstrate the effectiveness of our method. 
Response processes of $13$ Problem Solving in Technology-Rich Environments (PSTRE) items from $17$ countries are considered in this study. 
PIAAC was the first attempt to assess the PSTRE construct on a large scale and as a single dimension. Under the PIAAC framework, PSTRE is defined as the use of digital technology, communication tools, and the internet to obtain and evaluate information, communicate with others, and perform practical tasks \citep{organisation2012literacy}.
The survey also recorded a broad spectrum of respondents' background information such as gender, age, occupation, hourly income, education level, etc..
To conduct DIF analysis, we consider age, income, and gender as the demographic grouping variables. 
We include the process data of 8,398 respondents who answered all 13 items and have no missing value of the three covariates in the study. For age, we use $47$, which is the $70\%$ quantile, as the cutoff value to split the samples into younger and older sub-populations. The younger population is treated as the reference group, and the older population is treated as the focal group. For income, we first group the samples by their nationality, and then use the medium income of each nation as the cutoff value. The lower-income and higher-income groups are treated as the focal and the reference groups, respectively. For gender, female is treated as the focal group and male as the reference group. 

Table \ref{tab:items} provides a descriptive summary of the $13$ items, where $n$ is the number of total possible actions, $\bar{L}$ is the average process sequence length, and Correct $\%$ is the percentage receiving the full credit on each item. When solving for each item, the respondents are presented with one or more simulated informational and communicative (ICT) environments, such as an email inbox, a spreadsheet, a web browser, etc. For example, in item U01a, the respondents are presented with an email inbox interface and are asked to classify the email senders into `can come' and `cannot come' categories based on their email contents. To complete the task, the respondents need to conduct a sequence of clicking, dragging, or typing actions, which are recorded in the log files as process data. 

\begin{table}[bt!]
\footnotesize
    \centering 
    {\renewcommand{\arraystretch}{1.3}
    \begin{tabular}{l c c c c}
    \hline
    Item ID & Description & $n$ & $\bar{L}$ & Correct \% \\ [0.5ex] 
    \hline
    U01a & Party Invitations $-$ Can/Cannot Come & 51 & 17.2 & 59.8 \\ 
    U01b & Party Invitations $-$ Accommodations & 55 & 26.1 & 52.4 \\
    U02 & Meeting Rooms & 100 & 26.9 & 15.7 \\
    U03a & CD Tally & 67 & 9.0 & 42.3 \\
    U04a & Class Attendance & 926 & 39.2 & 15.3 \\ [1ex] 
    U06a & Sprained Ankle $-$ Site Evaluation Table & 30 & 9.5 & 26.2 \\ [1ex] 
    U06b & Sprained Ankle $-$ Reliable/Trustworthy Site & 26 & 15.0 & 50.8 \\ [1ex] 
    U07 & Digital Photography Book Purcha & 40 & 18.6 & 51.7 \\ [1ex] 
    U11b & Locate E-mail $-$ File 3 E-mails & 137 & 24.8 & 26.5 \\ [1ex] 
    U16 & Reply All & 886 & 32.4 & 63.9 \\ [1ex]
    U19a & Club Membership $-$ Member ID & 162 & 17.3 & 75.1 \\ [1ex] 
    U19b & Club Membership $-$ Eligibility for Club President & 450 & 20.7 & 52.5 \\ [1ex] 
    U23 & Lamp Return & 164 & 21.7 & 38.2 \\ [1ex] 
    \hline
\end{tabular}}
\caption{Summary statistics of $13$ PIAAC problem-solving items. Here $n$ is the number of total possible actions, $\bar{L}$ is the average process sequence length, and Correct $\%$ is the percentage of correct answers.} \label{tab:items} 
\end{table}

\begin{figure}[!bt]
    \centering
    \includegraphics[width=0.6\linewidth]{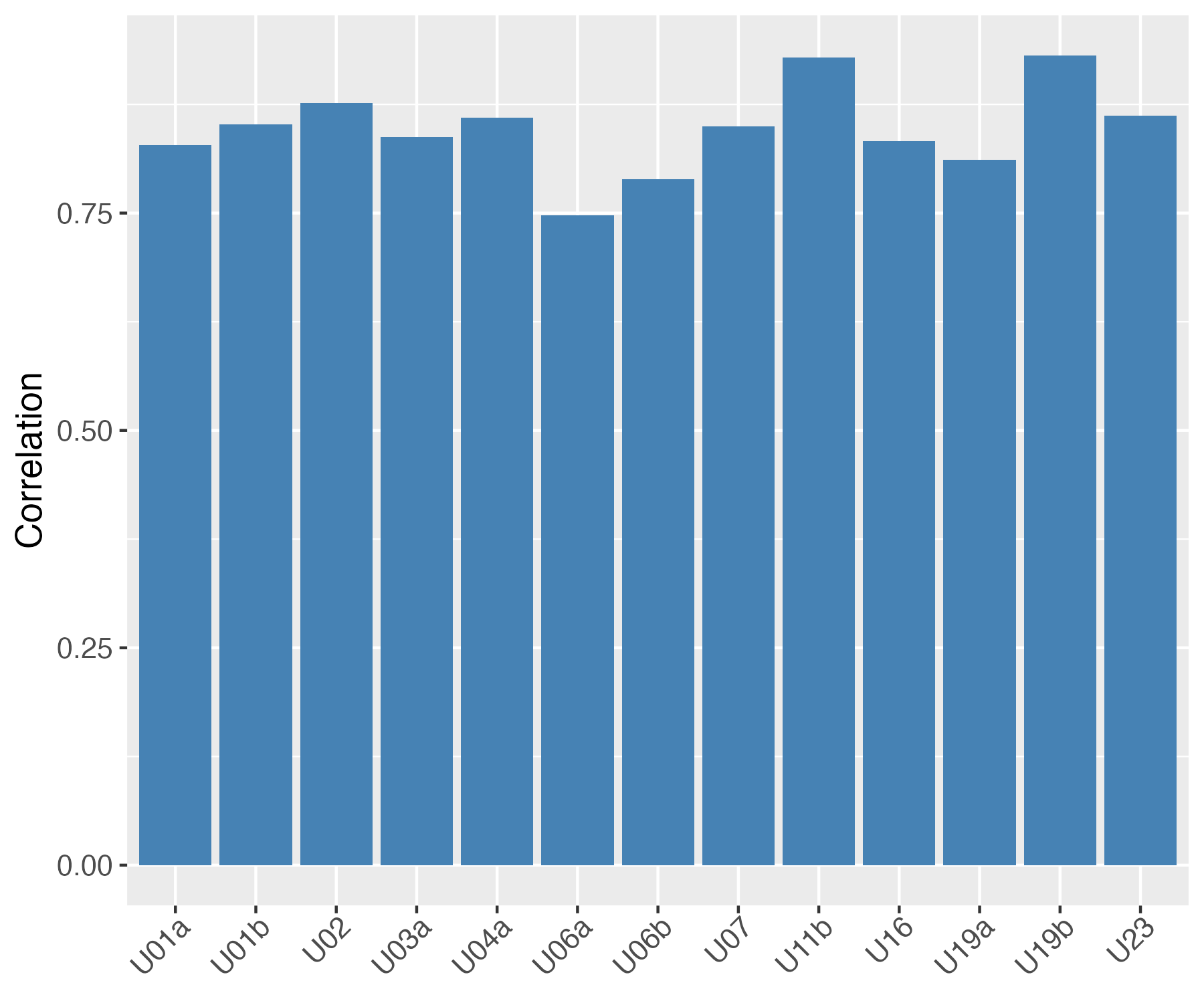}
    \caption{Out-of-sample prediction correlation of the process sequence length using the extracted features for each item.}
    \label{fig:len_pred}
\end{figure}

We use the MDS approach proposed in \cite{tang2020latent} to extract item features that approximate the geometric distances defined by a dissimilarity matrix of the action sequences. 
We set the dimension of features to be $K=100$ to ensure enough information is retained in the extracted features. 
To verify that the features contain an adequate amount of information in the process data, we use them to predict the responses with both linear regression and logistic regression. 
Results show that the extracted features can perfectly predict whether the examine received the full credit of each item with both regression methods. 
Engagement is often considered a nuisance trait in respondent behavior, and one plausible measure of engagement is the length of a respondent's process sequence. For each item, we randomly sample $80\%$ of training data to predict the process sequence length with the extracted features using ridge regression, where the ridge parameter is selected with cross-validation on the training data. Figure \ref{fig:len_pred} demonstrates the out-of-sample correlation between the predicted and actual process sequence lengths for each item. It shows that the extracted features are able to predict the sequence length with high accuracy. 

\subsection{DIF Existence}
The proposed procedure assumes that we have access to the ground truth target trait values. 
However, these target traits are unknown and need to be estimated in practice. 
To obtain initial estimates of the target traits, we utilize the responses to all $13$ items. We then use these target trait estimates to identify DIF items. 
For the linear model, the latent traits are estimated by performing maximum-likelihood factor analysis on the response data. 
For the M2PL model, the link function $g(\cdot)$ in Equation \eqref{eq:glm} is the logit function, and the reduced model without the nuisance trait is calibrated by maximizing the marginal likelihood using the expectation-maximization algorithm \citep{bock1981marginal}.
The initial estimation $\hat\bt^{(0)}$ is the MLE estimate after item calibration. 

Table \ref{tab:detection} summarizes the DIF detection results using the initial trait estimate $\hat{\bt}^{(0)}$. For the age variable, $12$ out of $13$ items are detected to have uniform DIF with both the linear model and the logistic model. For the income variable, $5$ items have uniform DIF with both models. For the gender variable, $5$ are uniform-DIF items with the linear model and $7$ with the logistic model.


\begin{table}[hbt!]
\centering
\footnotesize
\begin{tabular}{ccccccc}
\toprule
\multirow{4}{*}{Item} & \multicolumn{6}{c}{$\hat\lambda$ ($\hat\sigma(\hat\lambda)$)} \\
\cmidrule(lr){2-7}
 & \multicolumn{3}{c}{Linear model} & \multicolumn{3}{c}{Logistic model} \\
\cmidrule(lr){2-4} \cmidrule(lr){5-7}
 & Age & Income & Gender & Age & Income & Gender \\
\midrule
U01a & -$\mathbf{0.256 (0.018)}$ & $\mathbf{0.032 (0.016)}$ & -$\mathbf{0.040 (0.016)}$ & -$\mathbf{0.861 (0.067)}$ & $0.117 (0.064)$ & -$\mathbf{0.147 (0.063)}$ \\
U01b & -$\mathbf{0.074 (0.018)}$ & $0.012 (0.017)$ & -$0.005 (0.016)$ & -$\mathbf{0.238 (0.065)}$ & $0.023 (0.060)$ & $0.003 (0.059)$\\
U02 & $\mathbf{0.088 (0.021)}$ & -$\mathbf{0.041 (0.019)}$ & -$0.019 (0.019)$ & $\mathbf{0.377 (0.092)}$ & -$\mathbf{0.152 (0.076)}$ & $0.022 (0.076)$ \\
U03a & -$\mathbf{0.076 (0.020)}$ & -$0.020 (0.018)$ & -$\mathbf{0.093 (0.018)}$ & -$\mathbf{0.231 (0.063)}$ & -$0.067 (0.056)$ & -$\mathbf{0.266 (0.056)}$ \\
U04a & $\mathbf{0.089 (0.022)}$ & -$\mathbf{0.05 (0.020)}$ & -$0.032 (0.020) $& $\mathbf{0.345 (0.082)}$ & -$\mathbf{0.161 (0.069)}$ & -$0.054 (0.069)$\\
U06a & $\mathbf{0.123 (0.021)}$ & -$\mathbf{0.045 (0.019)}$ & $\mathbf{0.039 (0.019)}$ &$\mathbf{0.421 (0.068)}$ & -$\mathbf{0.133 (0.059)}$ & $\mathbf{0.172 (0.059)}$ \\
U06b & $\mathbf{0.063 (0.023)}$ & -$0.037 (0.021)$ & $0.028 (0.021)$ & $\mathbf{0.164 (0.052)}$ & -$0.082 (0.047)$ & 
$0.067 (0.046)$\\
U07 & $\mathbf{0.132 (0.020)}$ & -$0.023 (0.018)$ & $\mathbf{0.076 (0.018)}$ & $\mathbf{0.426 (0.061)}$ & -$0.080 (0.054)$ & $\mathbf{0.239 (0.054)}$ \\
U11b & -$\mathbf{0.051 (0.021)}$ & $0.023 (0.019)$ & $0.026 (0.019)$ & -$\mathbf{0.207 (0.070)}$ & $0.081 (0.059)$ & $\mathbf{0.134 (0.059)}$\\
U16 & -$\mathbf{0.037 (0.019)}$ & $\mathbf{0.053 (0.017)}$ & $\mathbf{0.052 (0.017)}$ & -$0.078 (0.065)$ & $\mathbf{0.164 (0.061)}$ & $\mathbf{0.176 (0.061)}$\\
U19a & $\mathbf{0.069 (0.020)}$ & -$0.001 (0.018)$ & $\mathbf{0.046 (0.018)}$ & $\mathbf{0.347 (0.070)}$ & -$0.025 (0.066)$ & $0.127 (0.065)$ \\
U19b & $\mathbf{0.074 (0.018)}$ & $0.012 (0.017)$ & -$\mathbf{0.076 (0.016)}$ & $\mathbf{0.269 (0.065)}$ & $0.016 (0.059)$ & -$\mathbf{0.257 (0.059)}$ \\
U23 & -$0.033 (0.020)$ & $\mathbf{0.051 (0.018)}$ & -$0.001 (0.018)$ & -$\mathbf{0.115 (0.064)}$ & $\mathbf{0.157 (0.056)}$ & $0.034 (0.056)$ \\
\bottomrule
\end{tabular}
\caption{DIF detection results \emph{without} the nuisance trait. Bold text indicates statistical significance under the $0.05$ significance level.}
\label{tab:detection}
\end{table}

\subsection{DIF Correction}
We implement the proposed method to estimate the nuisance traits with the extracted process features, and to correct for DIF effects. For the linear model, we use the closed-form expression in Proposition \ref{prop} as the estimate, while for the M2PL model, the nuisance traits are estimated by solving \eqref{eq:minimizer} using the \texttt{optim} function with the L-BFGS-B optimizer in \texttt{R}. 

\begin{figure}[bt!]
    \centering
    \includegraphics[width=0.9\textwidth]{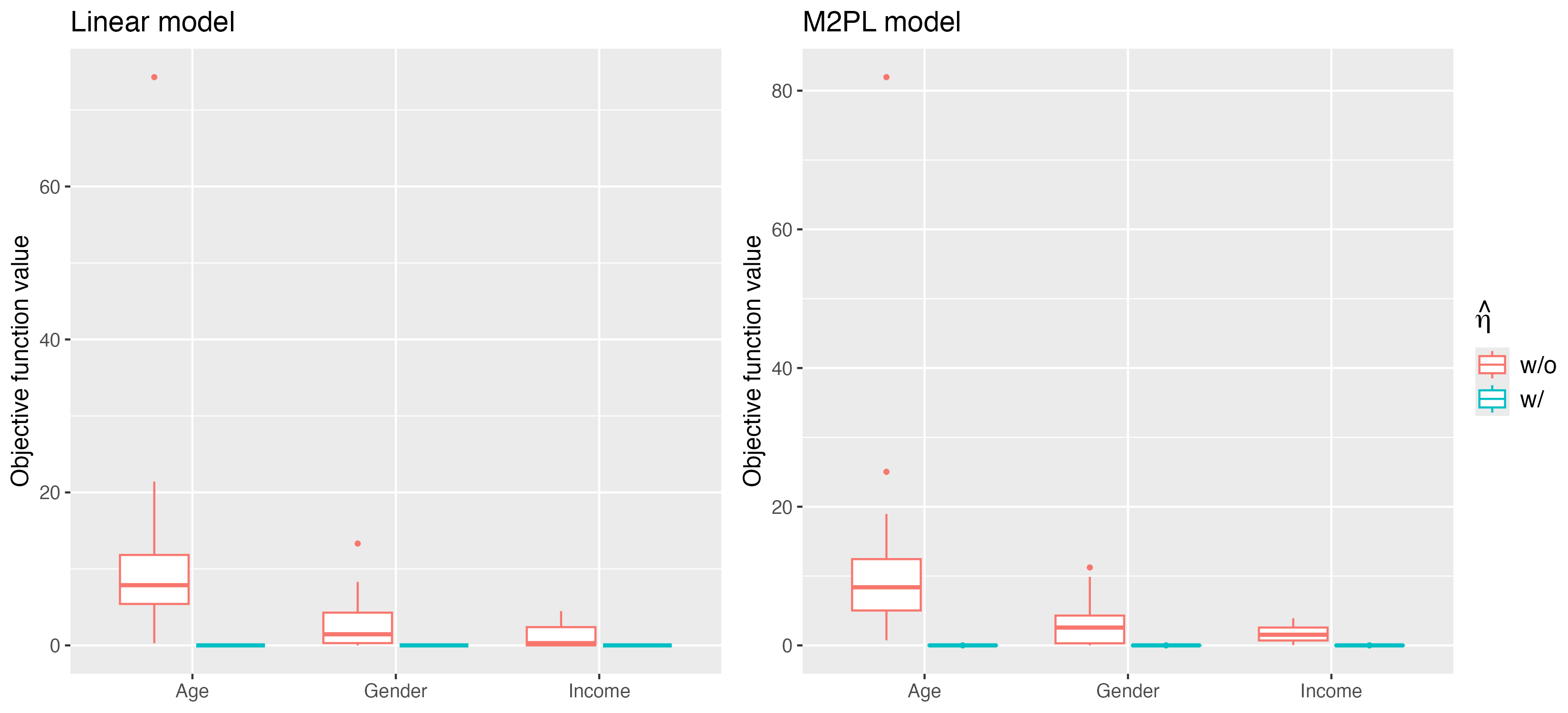}
    \caption{Comparing the objective function value with and without the nuisance trait surrogate for the linear model (left) and the M2PL model (right) with one grouping variable.}
    \label{fig:obj}
\end{figure}

Figure \ref{fig:obj} includes the boxplots of the objective function \eqref{eq:minimizer} with and without the nuisance trait surrogate for the three grouping variables. 
We see that the estimated nuisance traits serve the purpose of minimizing the objective functions for both models. 
Table \ref{tab:scores} demonstrates the sample mean Fisher information of the target trait $\theta$ before and after adding the nuisance trait surrogate $\hat{\bo\eta}$ for items that exhibit DIF. 
We see that the Fisher information increases in the linear model by adding the nuisance trait surrogate. For the M2PL model, we observe moderate reduction for most items, and significant reduction for items U06b and U07. For these two items, adding the nuisance trait surrogate significantly increases the the prediction accuracy of the item response.
The boxplots of the objective function with and without the nuisance trait when two grouping variables are present can be found in Figure \ref{fig:obj_2Z} in the Appendix. We see that the estimated nuisance traits are able to minimize the objective function with two grouping variables. 
For the case study, we do not have access to the ground truth target trait or a reliable unbiased estimator of the target trait as many items exhibit DIF. Therefore, we do not compare the between-group SS bias for the target trait estimates before and after DIF correction.

\begin{table}[!ht]
\footnotesize
\centering
\begin{tabular}{ccccccccc}
\toprule
\multirow{3}{*}{Item ID} 
 & \multicolumn{4}{c}{Linear model} & \multicolumn{4}{c}{M2PL model} \\
\cmidrule(lr){2-5} \cmidrule(lr){6-9}
 & w.o. $\hat\eta$ & Age & Income & Gender & w.o. $\hat\eta$ &  Age & Income & Gender \\
\midrule
U01a & 1.265 & 1.739 & 1.274 & 1.267 & 0.834 & 0.195 & -- & 0.569 \\
U01b & 0.896 & -- & -- & 0.897 & 0.796 & 0.335 & -- & -- \\
U02 & 1.183 & 1.231 & 1.294 & -- & 0.756 & 0.729 & 0.462 & -- \\
U03a & 0.623 & -- & 0.627 & 0.684 & 0.667 & 0.168 & -- & 0.176 \\
U04a & 0.312 & 0.337 & 0.322 & -- & 0.395 & 0.193 & 0.183 & -- \\
U06a & 0.345 & 0.397 & 0.361 & 0.351 & 0.489 & 0.146 & 0.285 & 0.248 \\
U06b & 0.133 & 0.137 & -- & -- & 0.146 & 0.030 & -- & 0.022 \\
U07 & 0.545 & 0.707 & -- & 0.610 & 0.507 & 0.007 & -- & 0.027 \\
U11b & 0.587 & 0.605 & -- & -- & 0.536 & 0.443 & -- & 0.273 \\
U16 & 0.822 & 0.827 & 0.851 & 0.840 & 0.645 & -- & 0.512 & 0.292 \\
U19a & 0.573 & 0.587 & -- & 0.584 & 0.507 & 0.340 & -- & 0.426 \\
U19b & 1.520 & -- & 1.523 & -- & 0.750 & 0.143 & 0.625 & -- \\
U23 & 0.867 & 0.870 & 0.912 & 0.867 & 0.619 & 0.540 & 0.358 & 0.312 \\
\bottomrule
\end{tabular}
\caption{Sample mean Fisher information for $\theta$ with and without nuisance trait surrogate, for three grouping variables and two models. Only the values corresponding to the DIF items are present.}
\label{tab:scores}
\end{table}

\begin{figure}[bt!]
    \centering
    \includegraphics[width=\linewidth]{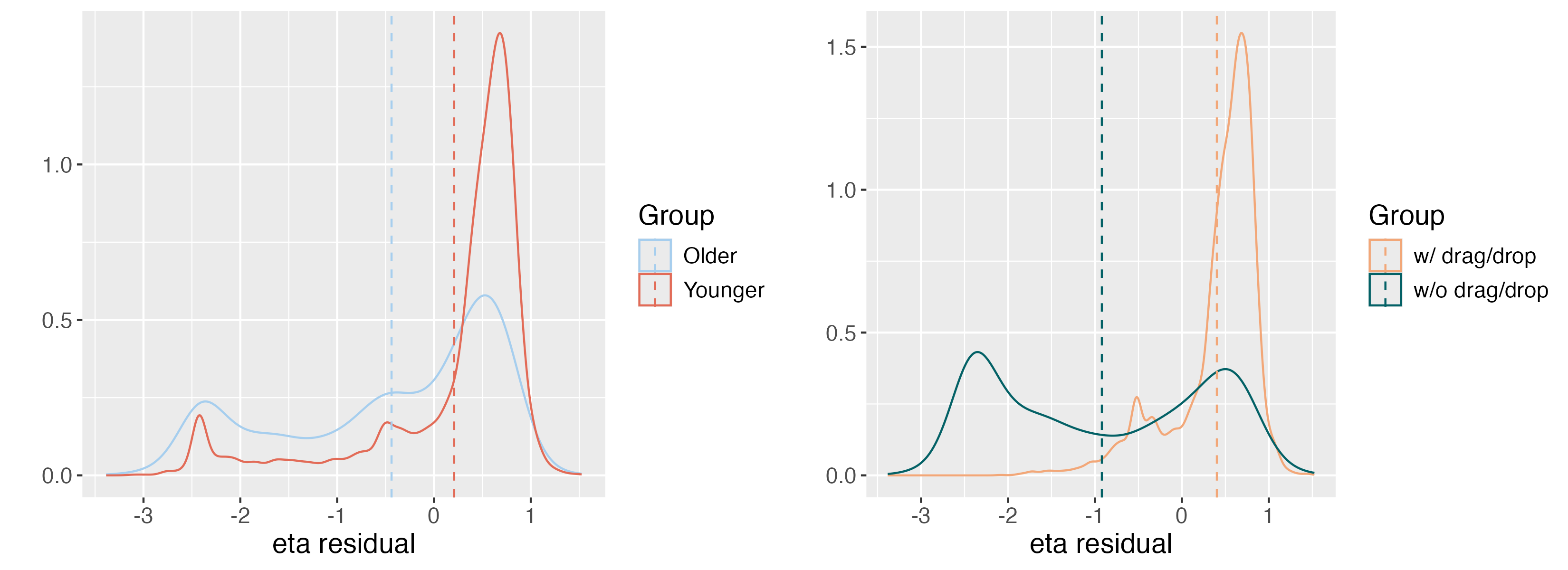}
    \caption{On the left: density plot of the residual nuisance trait $\tilde{\bo\eta}$ among the `old' group and the `young' group. On the right: density plot of the residual nuisance trait $\tilde{\bo\eta}$ among the group that used drag/drop actions and those that did not.}
    \label{fig:eta}
\end{figure}

As an illustration, we consider item U01a and the M2PL model to interpret the results for the age variable. 
After obtaining the estimated nuisance traits, we update the estimation of the target trait as $\hat{\bo\theta}$ by solving Eq \eqref{eq:update}. Recall that DIF arises when the functioning of the response differ given the latent trait. Therefore, we study the characteristics of the residual nuisance trait given the target trait, i.e., $\tilde{\bo\eta}=\hat{\bo\eta} - \mathbb{E}[\hat{\bo\eta}|\bo\theta]$. 
To interpret why DIF occurs in item U01a, we check the original process sequences corresponding to the minimum and maximum values of $\tilde{\bo\eta}$, and find that the usage of using drag/drop actions is related to the value of the residual nuisance trait.
To verify this assumption, we calculate the correlation between $\tilde{\bo\eta}$ and whether drag/drop actions are used, which achieves $0.61$.
And the correlation between $\tilde{\bo\eta}$ and the number of drag/drop actions achieves $0.49$.
These results suggest that the estimated nuisance variable can indicate the intensity of using drag/drop actions. Furthermore, the item response accuracy is $74.4\%$ among the group that used drag/drop actions, versus $26.3\%$ among those that did not use drag/drop actions.
Figure \ref{fig:eta} demonstrates the density plots of the residual nuisance trait $\tilde{\bo\eta}$ among the `older'/`younger' groups, and among the groups that did or did not use drag/drop actions. A possible interpretation to this phenomenon is that, more senior individuals might be less familiar with this type of drag-and-drop mouse usage. It is also possibly more error-prone for more senior individuals to move emails using drag-and-drop actions because of the small font size in the email interface and the narrow distances between email folders.

\section{Discussion}\label{sec: discussion}
Testing fairness is a prominent concern within psychometric and educational research, and DIF analysis is a commonly practiced approach to ensure testing fairness. 
When the distributions of the item response differ among two or more groups conditional on the target trait(s), DIF arises. 
Development of high-quality operational test items is costly, yet in practice, items identified with significant DIF effects are often discarded.   
In this paper, we propose a method to ``de-bias'' items that are detected with DIF by incorporating data beyond item responses. 
We utilize the rich information contained in item process data, which capture the whole response processes of respondents when they interact with computer-based items. 
Specifically, we attribute DIF to multidimensionality, where nuisance traits with heterogeneous sub-group distributions also affect the item responses, besides the target trait to be measured. 
To uncover the unobserved nuisance trait, we propose to minimize the maximum likelihood difference of the models with and without the grouping variable.
In the simple case with linear regression models and one grouping variable, there is a closed-form solution to the proposed optimization problem.
Simulation studies and a real data case study demonstrate the effectiveness of the proposed method.

Some limitations do exist in the current method.
Firstly, introducing the nuisance trait into the model might reduce measurement reliability
, as suggested in some decrease in the Fisher information for the target variable in the case study. 
It is of interest to study if a weighted summation of the maximum likelihood reduction as in \eqref{eq:objective} and model liability quantification such as the FI would be appropriate as the new objective function. 
Secondly, our proposed method relies on identifying a set of DIF-free or anchor items to identify the DIF items. 
In the case study, the initial targets are estimated assuming that all the items are DIF-free, and then utilized for DIF detection.
However, this approach might be prone to bias when the influence of DIF on the initial trait estimation is significant. 
In the future, we are interested in more sophisticated DIF detection methods such as item purification with stepwise model selection \citep{candell1988iterative, kopf2015anchor, kopf2015framework}.
In addition, we can bypass the tedious iterative purification procedure by employing methods similar to the covariate-adjusted model with regularization \citep{wang2023using, ouyang2024statistical}, where the anchor item identification and latent trait estimation are carried out simultaneously. However, model identifiability must be carefully examined, as existing methods in the literature cannot be directly applied to our setting.
Last but not least, process features are extracted from the action sequences and then utilized to linearly model the nuisance trait. However, a non-linear relationship between the nuisance trait and process data can be approximated by a neural network.

\clearpage
\bibliographystyle{apalike}
\bibliography{refs}

\clearpage
\appendix
\renewcommand{\theequation}{A\arabic{equation}}
\setcounter{equation}{0}
\renewcommand{\thesection}{\Alph{subsection}}
\setcounter{section}{0}

\section*{Appendix A: Mathematical Derivations}\label{sec:appendix-math}
\paragraph{Proof of Proposition \ref{prop}.}
Recall that $\Y^\dagger, \Z^\dagger, \X^\dagger$ are respectively the residuals of $\Y, \Z, \X$ after regressing on $(1,\bt)$.
Regress $\Z^\dagger$ on $\bo\eta=\X^\dagger\bw$:
\begin{equation*}
    \Z^\dagger = \beta \bo\eta + \bo\epsilon_Z,
\end{equation*}
and obtain the residuals $\hat{\bo\epsilon}_Z := \Z^\dagger - \hat{\beta}\bo\eta$ with $\hat\beta$ as the ordinary least squares (OLS) estimate. 
Specifically, since $\bo\eta^\top \bo\eta=1$ with $\X^{\dagger\top}\X^\dagger=\mathbf{I}_K$ and $\bw^\top\bw = 1$, we have
\begin{align}
    & \hat\beta =\frac{\Z^{\dagger\top}\bo\eta}{\bo\eta^\top \bo\eta}=\Z^{\dagger\top} \bo\eta, \notag \\
    & \hat{\bo\epsilon}_Z = \Z^\dagger - \bo\eta \Z^{\dagger\top}\bo\eta \label{eq:hat_eps_Z}
\end{align}
Similarly, obtain the OLS estimates for \eqref{eq:7}:
\begin{align}
    & \hat{a}'_1 = \frac{\Y^{\dagger\top} \bo\eta}{\bo\eta^\top \bo\eta} = \Y^{\dagger\top} \bo\eta, \notag \\
    & \hat{\bo\epsilon} = \Y^\dagger - \bo\eta \Y^{\dagger\top} \bo\eta \label{eq:hat_eps},
\end{align}
and the residuals $\hat{\bo\delta}=\Y^\dagger - \hat{a}_1 \bo\eta - \hat{\lambda} \Z^\dagger$ for \eqref{eq:8}.

It is straightforward to show that 
\begin{align*}
    \hat{\bo\epsilon} = \hat{\lambda} \hat{\bo\epsilon}_Z + \hat{\bo\delta}, \quad
    \hat{\lambda} = \frac{\hat{\bo\epsilon}^\top \hat{\bo\epsilon}_Z}{\|\hat{\bo\epsilon}_Z\|^2}.
\end{align*}
Therefore,
\begin{align*}
    \|\hat{\bo\epsilon}\|^2 - \|\hat{\bo\delta}\|^2 = \hat{\lambda}^2 \|\hat{\bo\epsilon}_Z \|^2= \frac{(\hat{\bo\epsilon}^\top \hat{\bo\epsilon}_Z)^2}{\|\hat{\bo\epsilon}_Z\|^2}.
\end{align*}
To obtain the zero of \eqref{eq:objective_linear} is equivalent to obtaining the zero of $\hat{\bo\epsilon}^\top \hat{\bo\epsilon}_Z$.
Denote $\hat\Y = \X^{\dagger\top} \Y^\dagger \in\mathbb{R}^{K}$ and $\hat\Z = \X^{\dagger\top} \Z^\dagger \in\mathbb{R}^K$. 
With $\bo\eta=\X^\dagger\bw,  \X^{\dagger\top}\X^\dagger=\mathbf{I}_K$, $\norm{\bo\eta}=\norm{\bw}=1$, plugging in \eqref{eq:hat_eps_Z}, \eqref{eq:hat_eps} yields
\begin{align*}
    \hat{\bo\epsilon}^\top \hat{\bo\epsilon}_Z & = (\Y^\dagger - \X^\dagger \bw \Y^{\dagger\top} \X^\dagger \bw )^\top (\Z^\dagger - \X^\dagger \bw \Z^{\dagger\top} \X^\dagger \bw) \\
    & = \Y^{\dagger\top} \Z^\dagger - \hat\Y^\top \bw \hat\Z^\top \bw - \bw^\top \hat\Y^\top \bw^\top \hat\Z^\top + \bw^\top \hat\Y \bw^\top \bw \hat\Z^\top \bw \\
    & = \Y^{\dagger\top} \Z^\dagger - \bw^\top \hat\Y \hat\Z^\top \bw\\
    & = \bw^\top \left(\Y^{\dagger\top} \Z^\dagger \cdot \mathbf{I}_K - \frac{\hat\Y \hat\Z^\top + \hat\Z \hat\Y^\top}{2}\right) \bw \\
    & := \bw^\top \A \bw,
\end{align*}
where $\A:= \Y^{\dagger\top} \Z^\dagger \cdot \mathbf{I}_K - \frac{\hat\Y \hat\Z^\top + \hat\Z \hat\Y^\top}{2}$.
In the above calculation, notice that $\bw^\top \hat\Y^\top$ and $\bw^\top \hat\Z^\top$ are numbers and therefore we could exchange order and take transpose when necessary. 
Consider the eigenvalue decomposition of $\A$: $\A = \Q \S \Q^\top$, where $\Q = (\bo q_1, \dots, \bo q_{K})\in\mathbb{R}^{K\times K}$, $\S=\diag (s_1, \dots, s_K)$, and $\Q^\top\Q=\mathbf{I}_K$. 
By the property of matrix $\A$, there are closed-form expressions of its two eigen vectors and $K$ eigenvalues. Specifically,
\begin{align*}
    \bo q_1 = \frac{\|\hat\Y\|\hat\Z + \|\hat\Z\|\hat\Y}{\left\|\|\hat\Y\|\hat\Z + \|\hat\Z\|\hat\Y \right\|} & , \  \bo q_2= \frac{\|\hat\Y\|\hat\Z - \|\hat\Z\|\hat\Y}{\left\| \|\hat\Y\|\hat\Z - \|\hat\Z\|\hat\Y \right\|}, \\
    s_1 = \Y^{\dagger\top} \Z^\dagger - \frac{\|\hat\Y\|\|\hat\Z\| + \hat\Y^\top\hat\Z}{2} & , \  s_2 = \Y^{\dagger\top} \Z^\dagger + \frac{\|\hat\Y\|\|\hat\Z\| - \hat\Y^\top\hat\Z}{2}, \\
    s_3=\dots & =s_K = \Y^{\dagger\top} \Z^\dagger.
\end{align*}
Under assumption \eqref{eq:condition}, we have $s_1 < 0 <s_3=\dots=s_K<s_2$.
Let
\begin{align*}
    \alpha = \sqrt{\frac{s_2}{s_2-s_1}}, \ \  
    \beta = \sqrt{\frac{-s_1}{s_2-s_1}} 
\end{align*}
and 
$$\hat\bw = \alpha\bo q_1 + \beta\bo q_2 \text{  or  } \alpha\bo q_1 - \beta\bo q_2,$$
then we have $\norm{\hat\bw}=1$ and $L(\hat\bw)=0$.

We notice that $L(\bw)$ has more than one zeros. We choose $\hat\bw_1 := \alpha \bo q_1 + \beta \bo q_2$ over $\hat\bw_2 := \alpha \bo q_1 - \beta \bo q_2$ because of the following geometric interpretations. We further write the scaled versions of $\hat\Y, \hat\Z$ as $\tilde\Y=\hat\Y / \|\hat\Y\|, \tilde\Z=\hat\Z / \|\hat\Z\|$. 
Without loss of generality, we have assumed that $\Y^{\dagger\top}\Z^\dagger>0$.
We consider the case where $\X^\dagger$ can almost perfectly predict $\Y^\dagger$. Then $\X^\dagger \X^{\dagger\top}\Y^\dagger \approx \Y^\dagger$, which leads to 
\begin{equation}\label{eq:angle}
    \hat\Y^\top \hat\Z = \Y^{\dagger\top} \X^\dagger \X^{\dagger\top} \Z^\dagger \approx \Y^{\dagger\top}\Z^\dagger>0.
\end{equation}
Denote the angle between $\hat\Y, \hat\Z$ as $\Delta$, which is also approximately the angle between $\Y^\dagger, \Z^\dagger$. By assumption, $\Delta\in (0, \pi/2)$.

\begin{figure}[hbt!]
    \centering
    \includegraphics[width=0.5\linewidth]{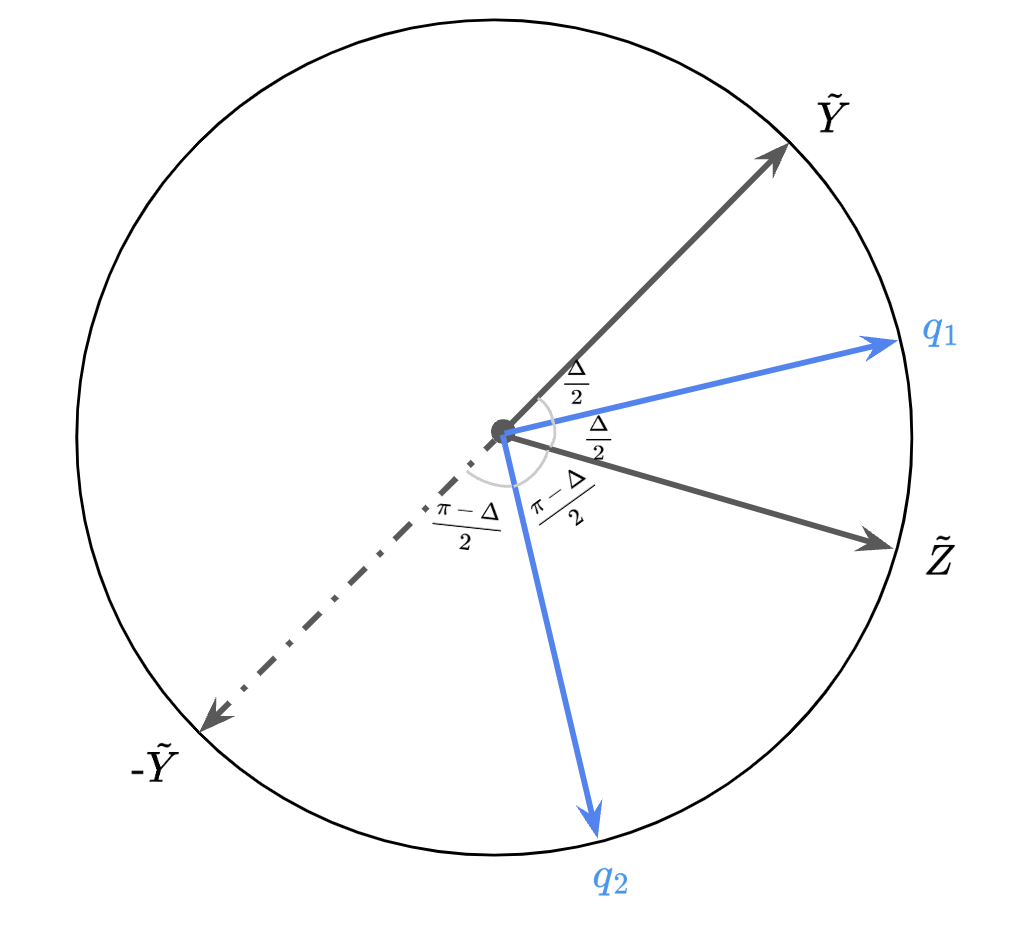}
    \caption{Geometric illustration for the proof of Proposition \ref{prop}.}
    \label{fig:angle}
\end{figure}

Rewrite $\bo q_1, \bo q_2$ as 
\begin{equation*}
    \bo q_1 = \frac{\tilde\Y + \tilde\Z}{\norm{\tilde\Y + \tilde\Z}}, \quad \bo q_2 = \frac{\tilde\Y - \tilde\Z}{\norm{\tilde\Y - \tilde\Z}}.
\end{equation*}
By \eqref{eq:angle}, we also notice that
\begin{align*}
    \frac{-s_1}{\|\hat\Y\| \|\hat\Z\|} \approx \frac{-\hat\Y^\top \hat\Z}{\|\hat\Y\| \|\hat\Z\|} + \frac{1}{2} + \frac{-\hat\Y^\top \hat\Z}{2\|\hat\Y\| \|\hat\Z\|} = \frac{1-\cos(\Delta)}{2} = \sin^2(\frac{\Delta}{2}),
\end{align*}
\begin{align*}
    \frac{s_2}{\|\hat\Y\| \|\hat\Z\|} \approx \frac{\hat\Y^\top \hat\Z}{\|\hat\Y\| \|\hat\Z\|} + \frac{1}{2} - \frac{-\hat\Y^\top \hat\Z}{2\|\hat\Y\| \|\hat\Z\|} = \frac{1+\cos(\Delta)}{2} = \cos^2(\frac{\Delta}{2}).
\end{align*}
Therefore, 
\begin{align*}
    \hat\bw_1 = \alpha \bo q_1 + \beta \bo q_2 \propto \cos(\frac{\Delta}{2}) \bo q_1 + \sin(\frac{\Delta}{2}) \bo q_2, \\
    \hat\bw_2 = \alpha \bo q_1 - \beta \bo q_2 \propto \cos(\frac{\Delta}{2}) \bo q_1 - \sin(\frac{\Delta}{2}) \bo q_2.
\end{align*}
Therefore, $\hat\bw_1$ aligns with $\tilde\Z$ and $\hat\bw_2$ aligns with $\tilde\Y$; see Figure \ref{fig:angle} for geometric interpretations. Furthermore, $\X^\dagger \hat\bw_1$ aligns with $\X^\dagger\X^{\dagger\top}\Z^\dagger$, which is the projection of $\Z^\dagger$ onto the column space of $\X^\dagger$. Similarly, $\X^\dagger \hat\bw_2$ aligns with the prediction of $\Y^\dagger$ by $\X^\dagger$. We choose to select $\hat\bw_1$ as the goal is to reduce DIF instead of using the process features to predict the response.

\qed

\clearpage
\section*{Appendix B: Additional Tables and Figures}

\subsection*{B.1. Simulation with Uniform DIF}

\begin{figure}[h!]
    \centering
    \includegraphics[width=0.65\linewidth]{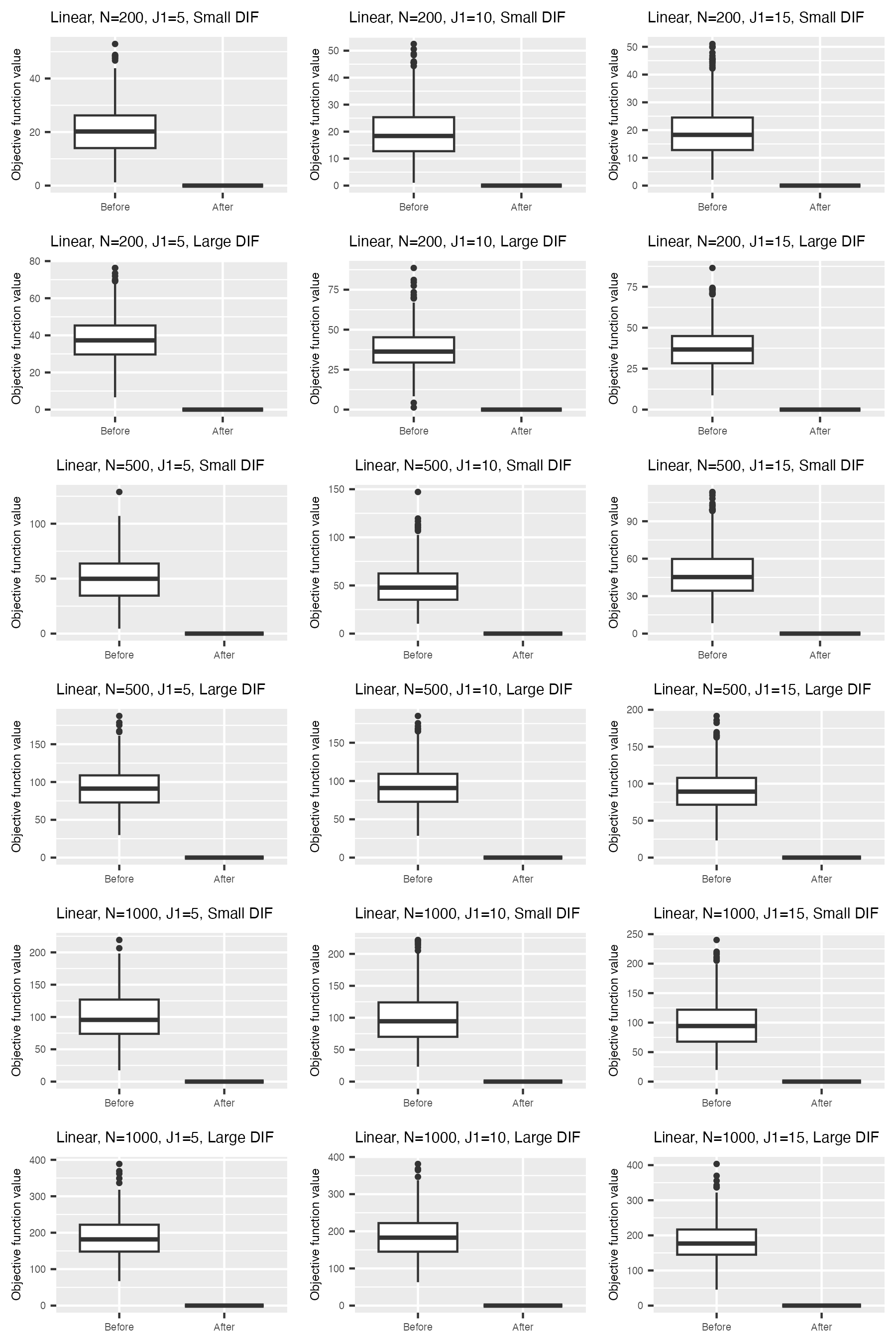}
    \caption{Objective function value before and after adding the nuisance trait surrogate for the \emph{linear model} with \emph{uniform} DIF, under different simulation settings.}
    \label{fig:sim_obj_linear}
\end{figure}

\begin{figure}
    \centering
    \includegraphics[width=0.65\linewidth]{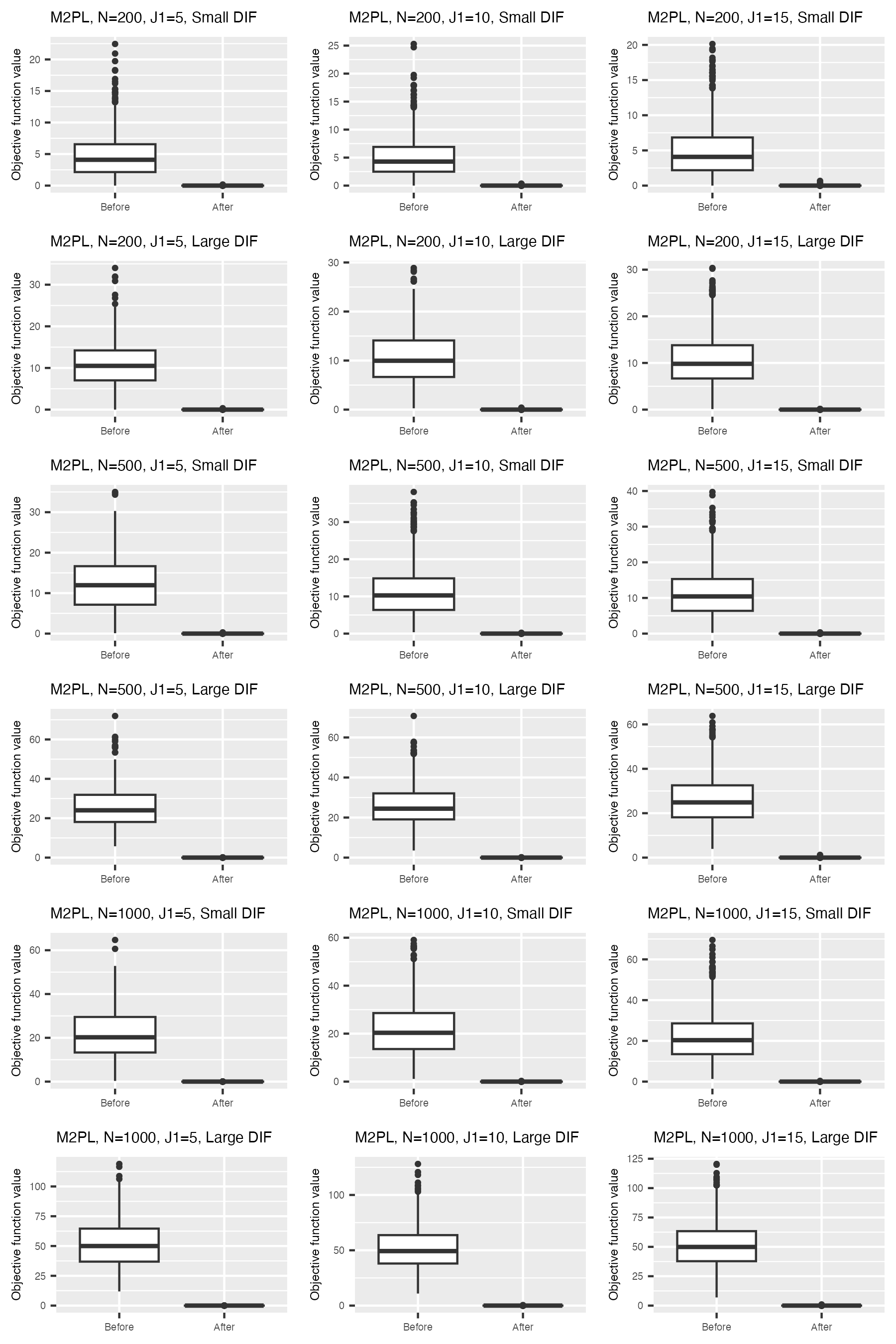}
    \caption{Objective function value before and after adding the nuisance trait surrogate for the \emph{M2PL model} with \emph{uniform} DIF, under different simulation settings.}
    \label{fig:sim_obj_m2pl}
\end{figure}

\begin{figure}
    \centering
    \includegraphics[width=0.65\linewidth]{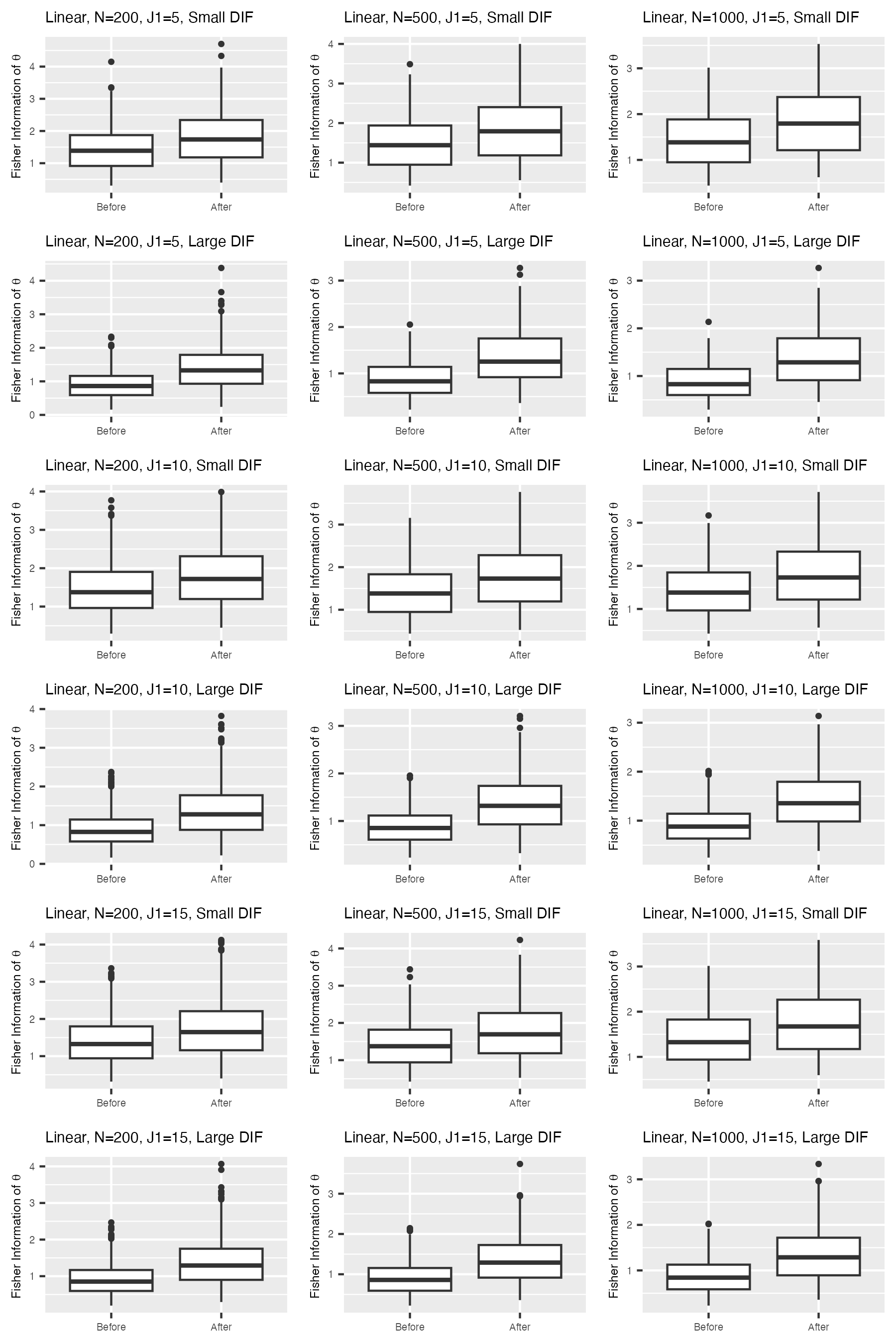}
    \caption{Fisher information of $\theta$ in the measurement model before and after adding the nuisance trait surrogate for the \emph{linear} model with \emph{uniform} DIF, under different simulation settings.}
    \label{fig:sim_info_linear_uniform}
\end{figure}

\begin{figure}
    \centering
    \includegraphics[width=0.65\linewidth]{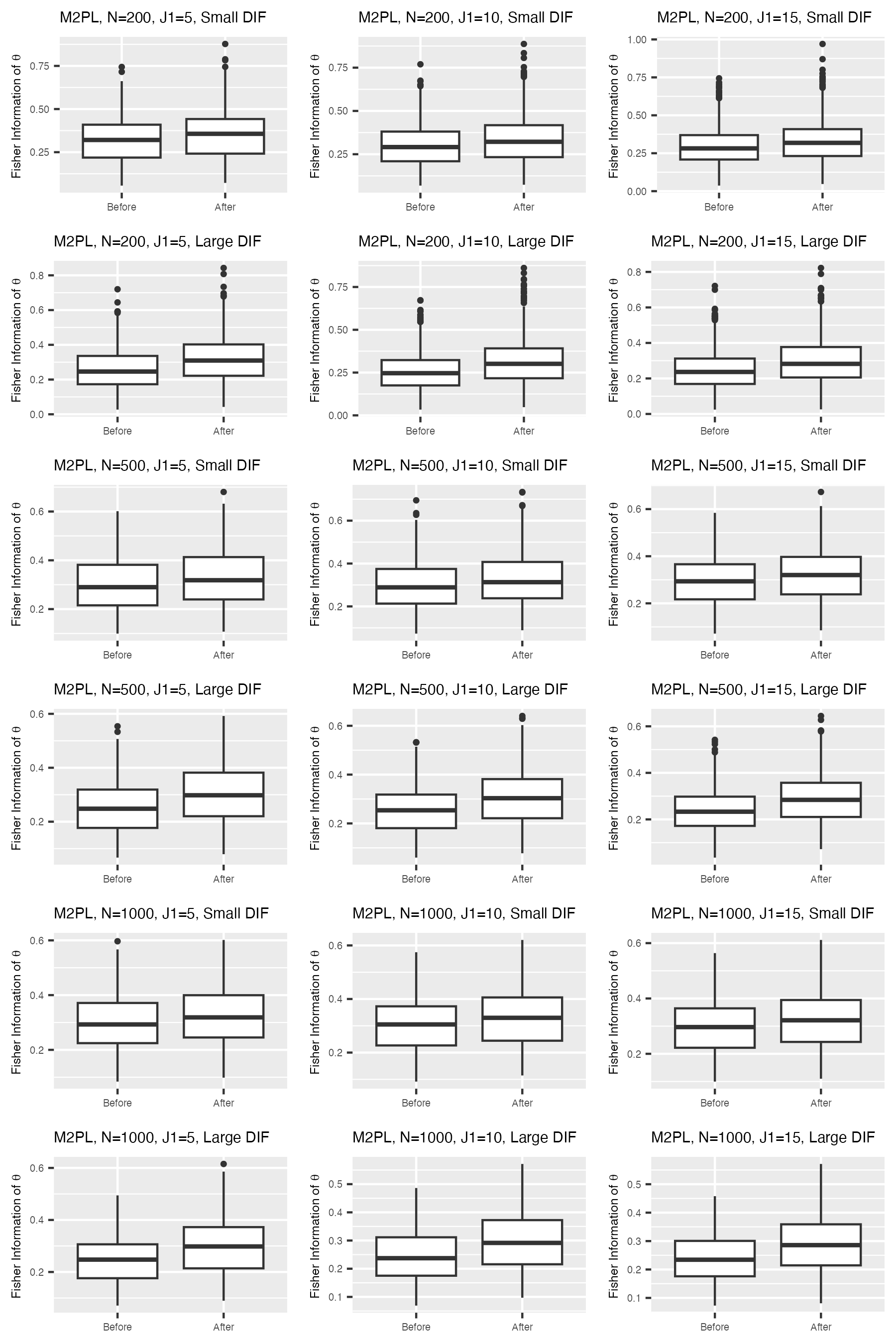}
    \caption{Fisher information of $\theta$ in the measurement model before and after adding the nuisance trait surrogate for the \emph{M2PL} model with \emph{uniform} DIF, under different simulation settings.}
    \label{fig:sim_info_M2PL_uniform}
\end{figure}

\begin{figure}
    \centering
    \includegraphics[width=0.63\linewidth]{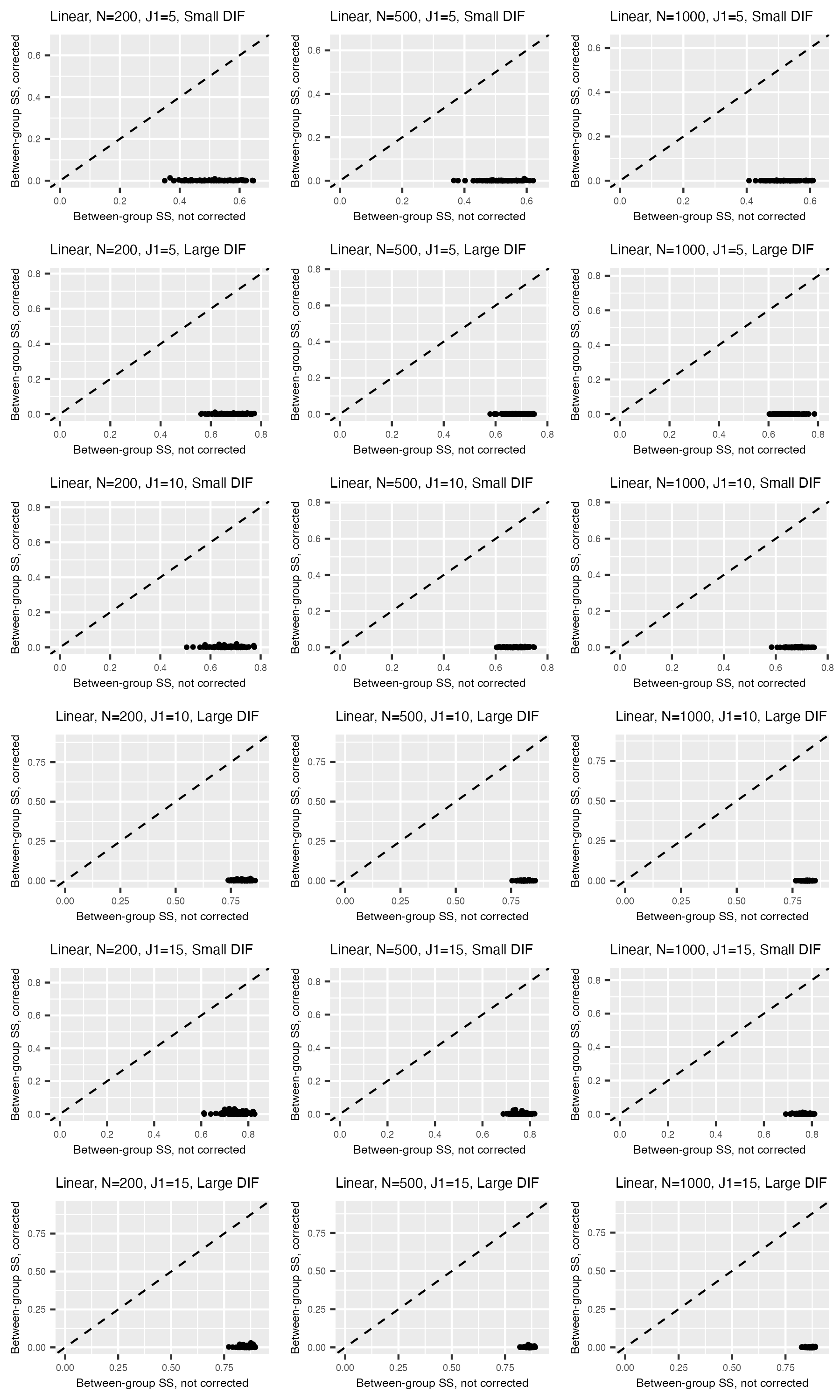}
    \caption{Between-group sum of squared bias for target trait estimation for the \emph{linear} model with \emph{uniform} DIF, under different simulation settings. The x-axis corresponds to the estimation without DIF correction using the DIF items; the y-axis corresponds to the DIF-corrected estimation using the DIF items.}
    \label{fig:sim_anova_linear_uniform}
\end{figure}

\begin{figure}
    \centering
    \includegraphics[width=0.63\linewidth]{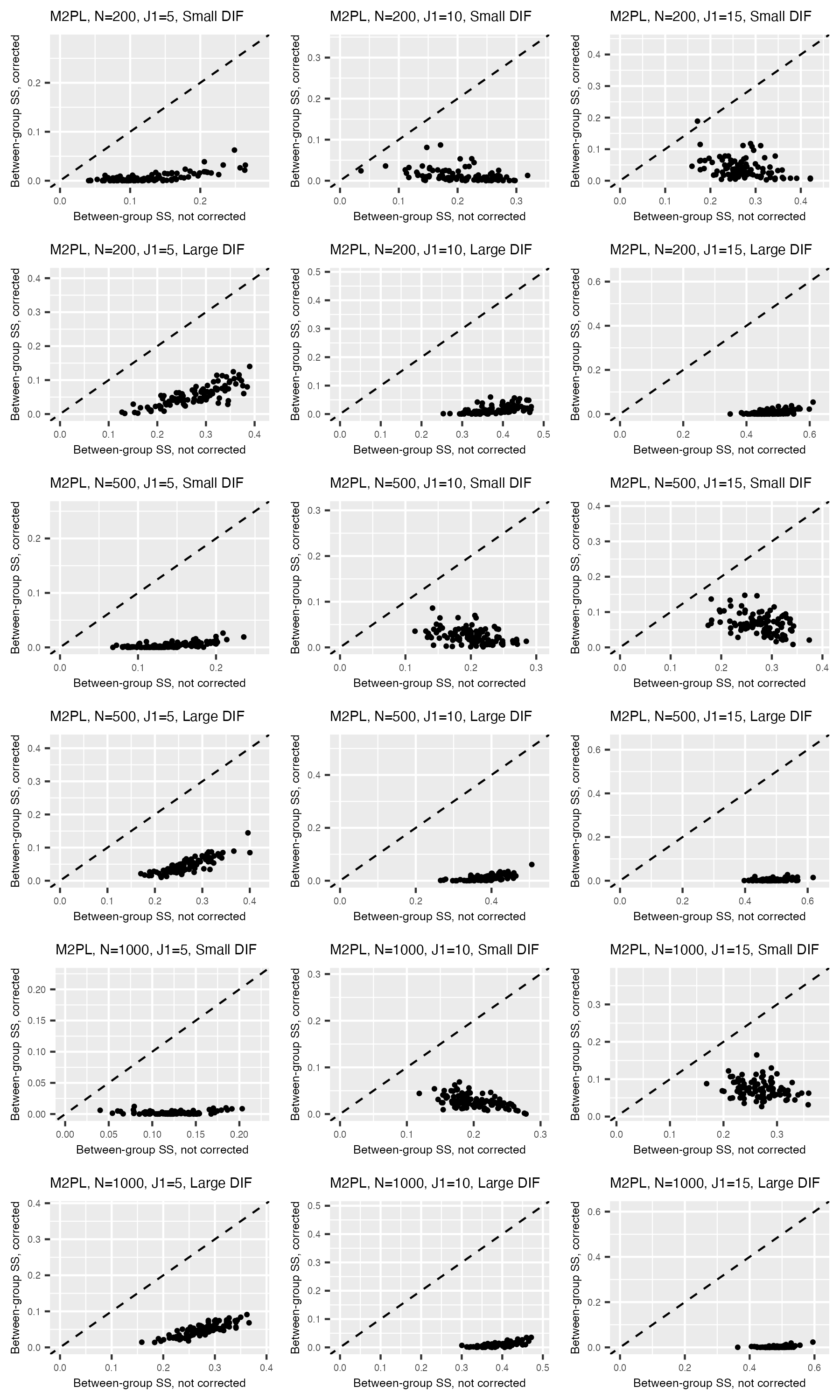}
    \caption{Between-group sum of squared bias for target trait estimation for the \emph{M2PL} model with \emph{uniform} DIF, under different simulation settings. The x-axis corresponds to the estimation without DIF correction using the DIF items; the y-axis corresponds to the DIF-corrected estimation using the DIF items.}
    \label{fig:sim_anova_M2PL_uniform}
\end{figure}

\clearpage
\subsection*{B.2. Simulation with Non-uniform DIF}
\begin{table}[hbt!]
\centering
\begin{tabular}{cccccccc}
\hline
& & \multicolumn{3}{c}{Small DIF} & \multicolumn{3}{c}{Large DIF} \\ \cmidrule(lr){3-5} \cmidrule(lr){6-8}
\multicolumn{2}{c}{} & $J_1$=5 & $J_1$=10 & $J_1$=15 & $J_1$=5 & $J_1$=10 & $J_1$=15 \\ 
\hline
\multirow{3}{*}{MSE ($d$)} & $N=200$ & 0.0484 & 0.0524 & 0.0410 & 0.0956 & 0.1206 & 0.1127 \\
 & $N=500$ & 0.0365 & 0.0389 & 0.0371 & 0.0955 & 0.0952 & 0.0957 \\
 & $N=1000$ & 0.0364 & 0.0355 & 0.0340 & 0.0881 & 0.0935 & 0.0959 \\ \hline
\multirow{3}{*}{MSE ($a_0$)} & $N=200$ & 0.0349 & 0.0319 & 0.0364 & 0.0916 & 0.0803 & 0.0815 \\
 & $N=500$ & 0.0300 & 0.0253 & 0.0268 & 0.0713 & 0.0623 & 0.0722 \\
 & $N=1000$ & 0.0248 & 0.0270 & 0.0238 & 0.0647 & 0.0640 & 0.0627 \\ \hline
\multirow{3}{*}{MSE ($a_1$)} & $N=200$ & 0.0538 & 0.0582 & 0.0586 & 0.1494 & 0.1513 & 0.1505 \\
 & $N=500$ & 0.0508 & 0.0543 & 0.0539 & 0.1325 & 0.1301 & 0.1400 \\
 & $N=1000$ & 0.0499 & 0.0532 & 0.0514 & 0.1461 & 0.1400 & 0.1312 \\ \hline
 \multirow{3}{*}{Corr ($\hat\be, \be$)} & $N=200$ & 0.7264 & 0.7238 & 0.7198 & 0.7151 & 0.7177 & 0.7200 \\
& $N=500$ & 0.7254 & 0.7201 & 0.7229 & 0.7306 & 0.7319 & 0.7243 \\
 & $N=1000$ & 0.7241 & 0.7197 & 0.7230 & 0.7193 & 0.7226 & 0.7291 \\ \hline
\end{tabular}
\caption{Mean squared error of item parameter estimates and nuisance trait correlation for the \emph{linear model} with \emph{non-uniform} DIF under different simulation settings. The values are averaged across the DIF items and replications.}
\label{tab:mse_linear_nonuniform}
\end{table}

\begin{table}[hbt!]
\centering
\begin{tabular}{cccccccc}
\hline
& & \multicolumn{3}{c}{Small DIF} & \multicolumn{3}{c}{Large DIF} \\ \cmidrule(lr){3-5} \cmidrule(lr){6-8}
\multicolumn{2}{c}{} & $J_1$=5 & $J_1$=10 & $J_1$=15 & $J_1$=5 & $J_1$=10 & $J_1$=15 \\ 
\hline
\multirow{3}{*}{MSE ($d$)} & $N=200$ & 0.0602 & 0.0564 & 0.0527 & 0.0613 & 0.0568 & 0.0631 \\
 & $N=500$ & 0.0223 & 0.0215 & 0.0208 & 0.0233 & 0.0215 & 0.0242 \\
 & $N=1000$ & 0.0107 & 0.0111 & 0.0119 & 0.0144 & 0.0129 & 0.0159 \\ \hline
\multirow{3}{*}{MSE ($a_0$)} & $N=200$ & 0.1068 & 0.0947 & 0.0982 & 0.1006 & 0.1086 & 0.1133 \\
 & $N=500$ & 0.0440 & 0.0496 & 0.0602 & 0.0540 & 0.0643 & 0.0677 \\
 & $N=1000$ & 0.0307 & 0.0347 & 0.0498 & 0.0430 & 0.0490 & 0.0566 \\ \hline
\multirow{3}{*}{MSE ($a_1$)} & $N=200$ & 0.0633 & 0.0737 & 0.0663 & 0.1115 & 0.1009 & 0.1226 \\
 & $N=500$ & 0.0303 & 0.0301 & 0.0323 & 0.0724 & 0.0768 & 0.0876 \\
 & $N=1000$ & 0.0221 & 0.0252 & 0.0257 & 0.0704 & 0.0725 & 0.0827 \\ \hline
 \multirow{3}{*}{Corr ($\hat\be, \be$)} & $N=200$ & 0.7245 & 0.7005 & 0.7002 & 0.8052 & 0.8118 & 0.7981 \\
& $N=500$ & 0.8052 & 0.7958 & 0.7887 & 0.8580 & 0.8531 & 0.8457 \\
 & $N=1000$ & 0.8410 & 0.8319 & 0.8263 & 0.8771 & 0.8759 & 0.8655 \\ \hline
\end{tabular}
\caption{Mean squared error of item parameter estimates and nuisance trait correlation for the \emph{M2PL model} with \emph{non-uniform} DIF under different simulation settings. The values are averaged across the DIF items and replications.}
\label{tab:mse_m2pl_nonuniform}
\end{table}

\begin{figure}[hbt!]
    \centering
    \includegraphics[width=0.7\linewidth]{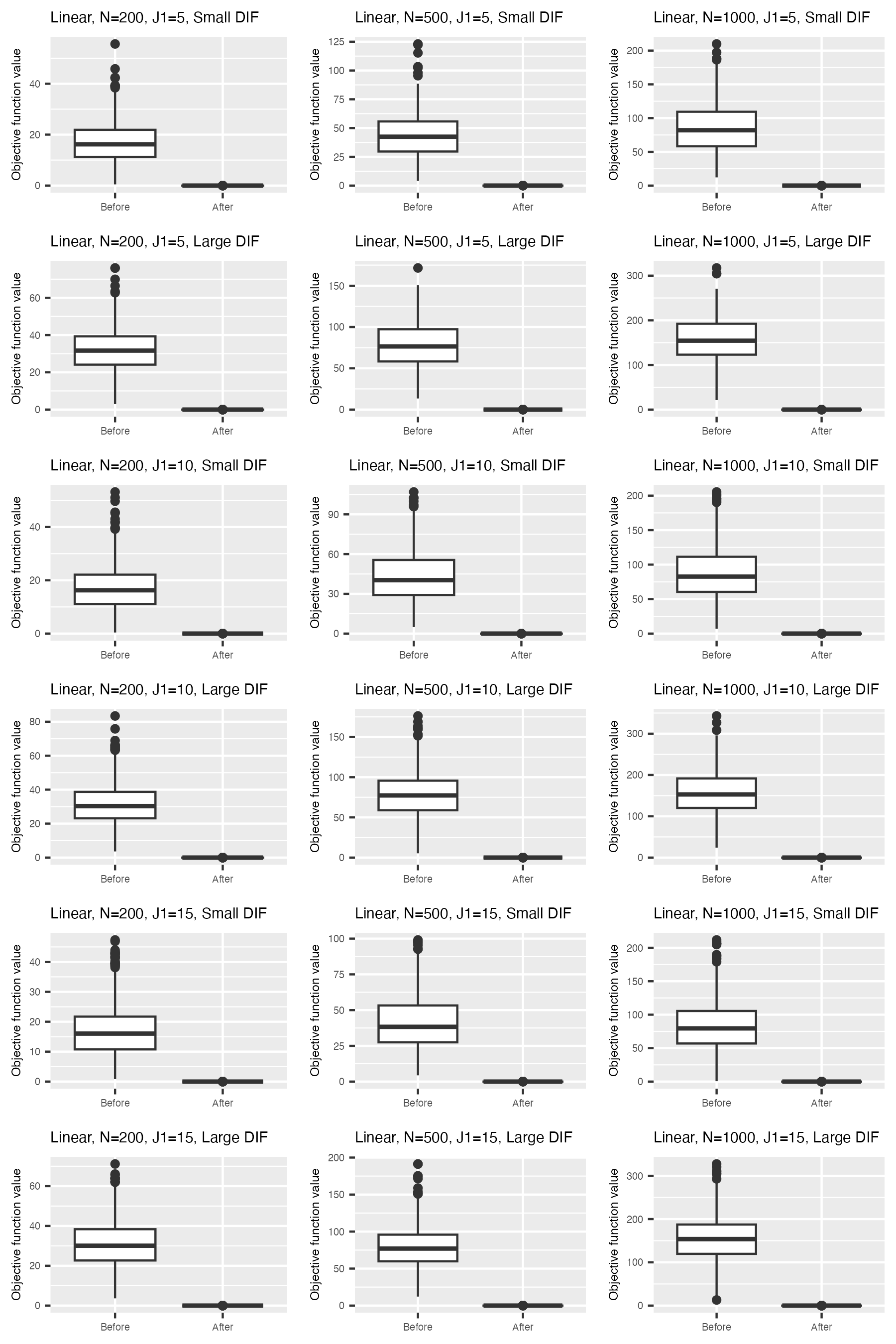}
    \caption{Objective function value before and after adding the nuisance trait surrogate for the \emph{linear model} with \emph{non-uniform} DIF, under different simulation settings.}
    \label{fig:sim_obj_linear_nonuniform}
\end{figure}

\begin{figure}[hbt!]
    \centering
    \includegraphics[width=0.7\linewidth]{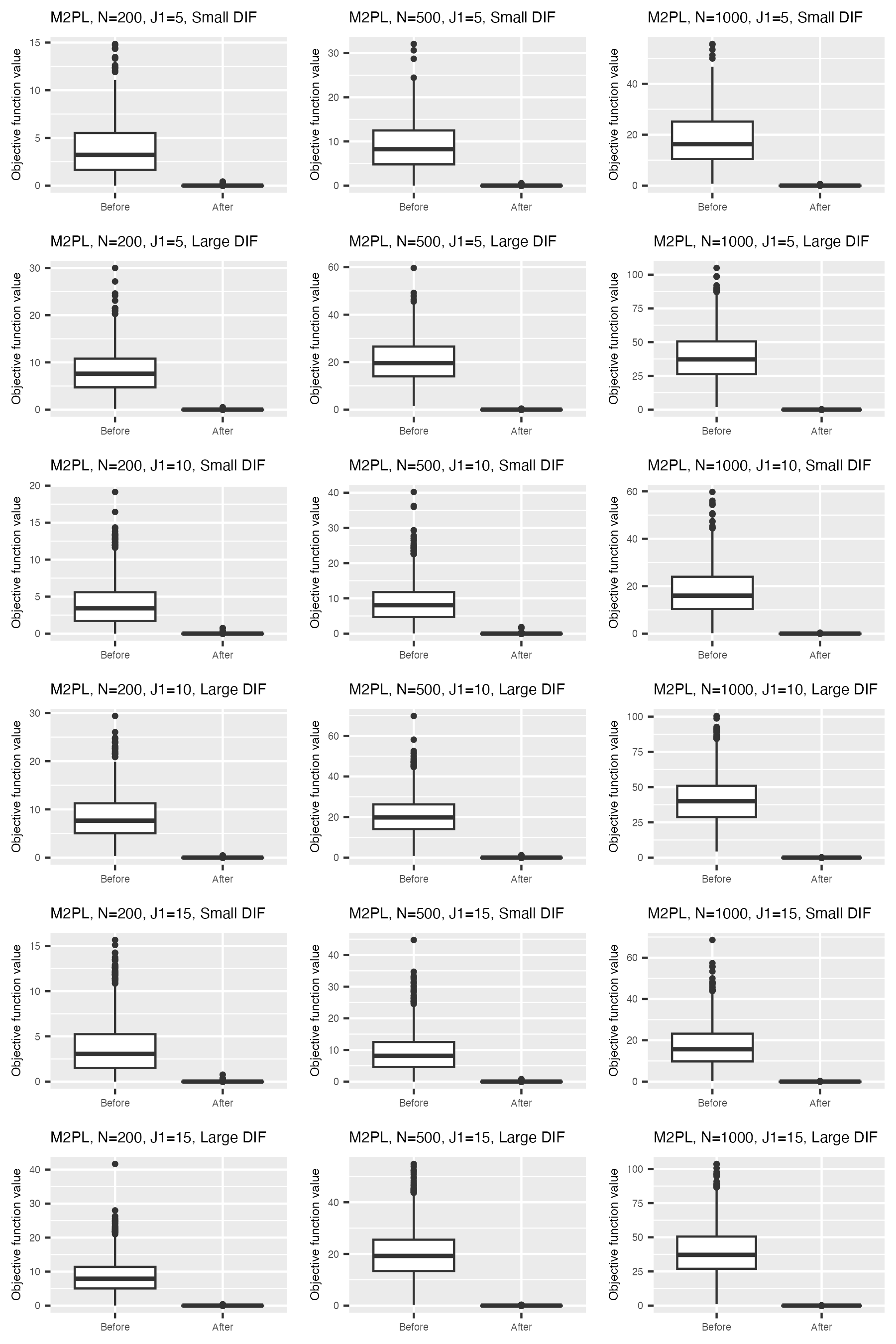}
    \caption{Objective function value before and after adding the nuisance trait surrogate for the \emph{M2PL model} with \emph{non-uniform} DIF, under different simulation settings.}
    \label{fig:sim_obj_m2pl_nonuniform}
\end{figure}

\begin{figure}
    \centering
    \includegraphics[width=0.7\linewidth]{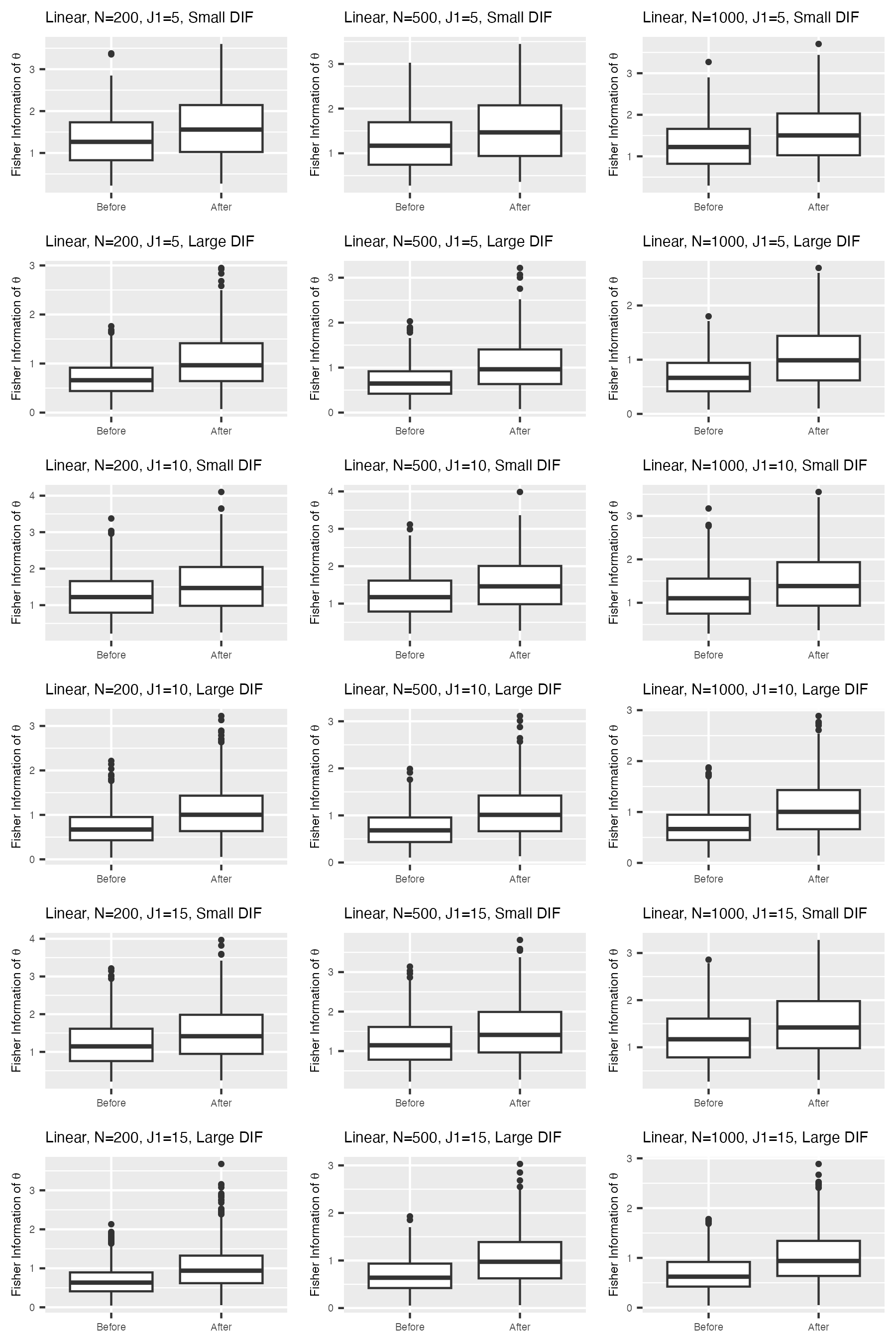}
    \caption{Fisher information of $\theta$ in the measurement model before and after adding the nuisance trait surrogate for the \emph{linear} model with \emph{non-uniform} DIF, under different simulation settings.}
    \label{fig:sim_info_linear_nonuniform}
\end{figure}

\begin{figure}
    \centering
    \includegraphics[width=0.7\linewidth]{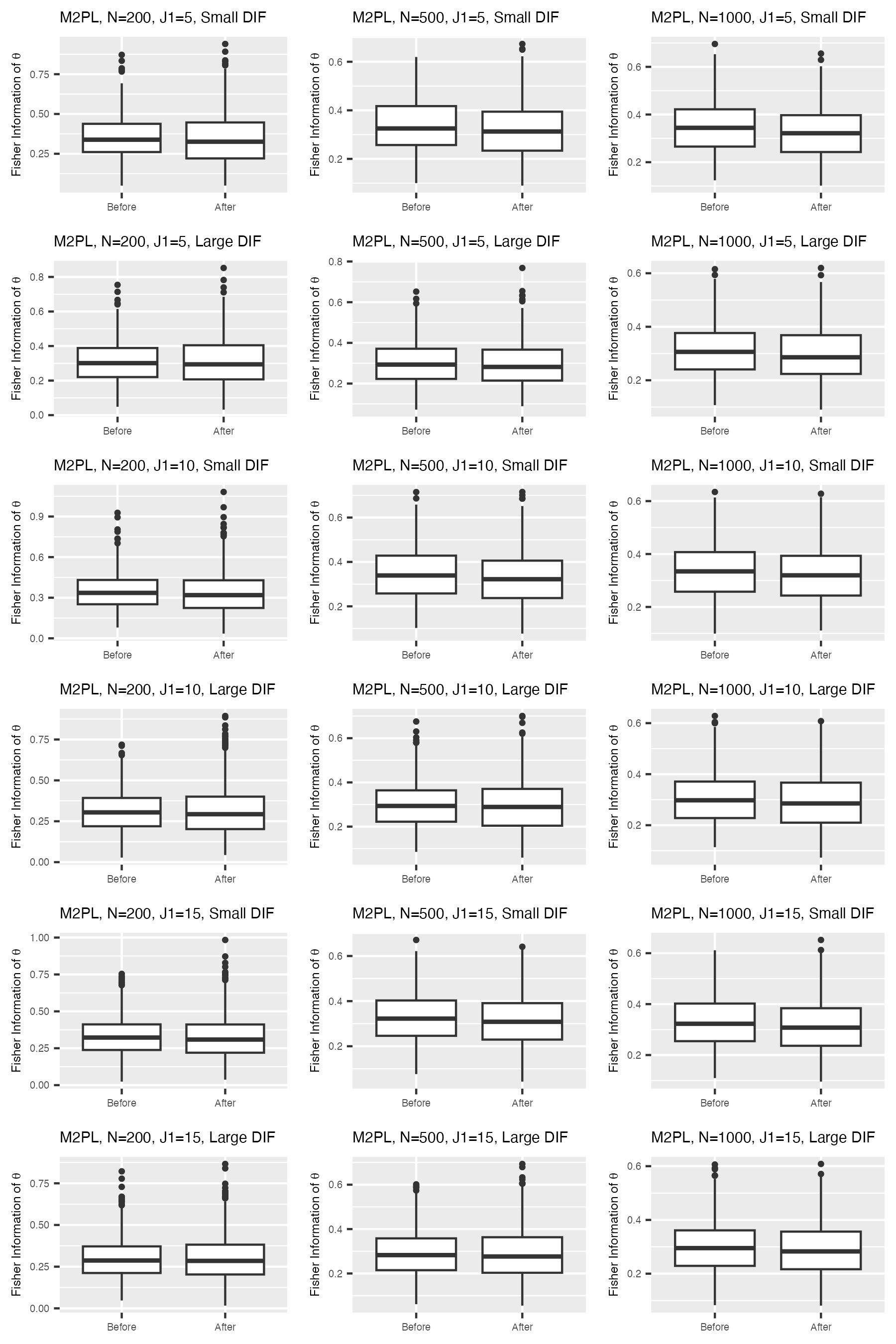}
    \caption{Fisher information of $\theta$ in the measurement model before and after adding the nuisance trait surrogate for the \emph{M2PL} model with \emph{non-uniform} DIF, under different simulation settings.}
    \label{fig:sim_info_m2pl_nonuniform}
\end{figure}

\begin{figure}
    \centering
    \includegraphics[width=0.63\linewidth]{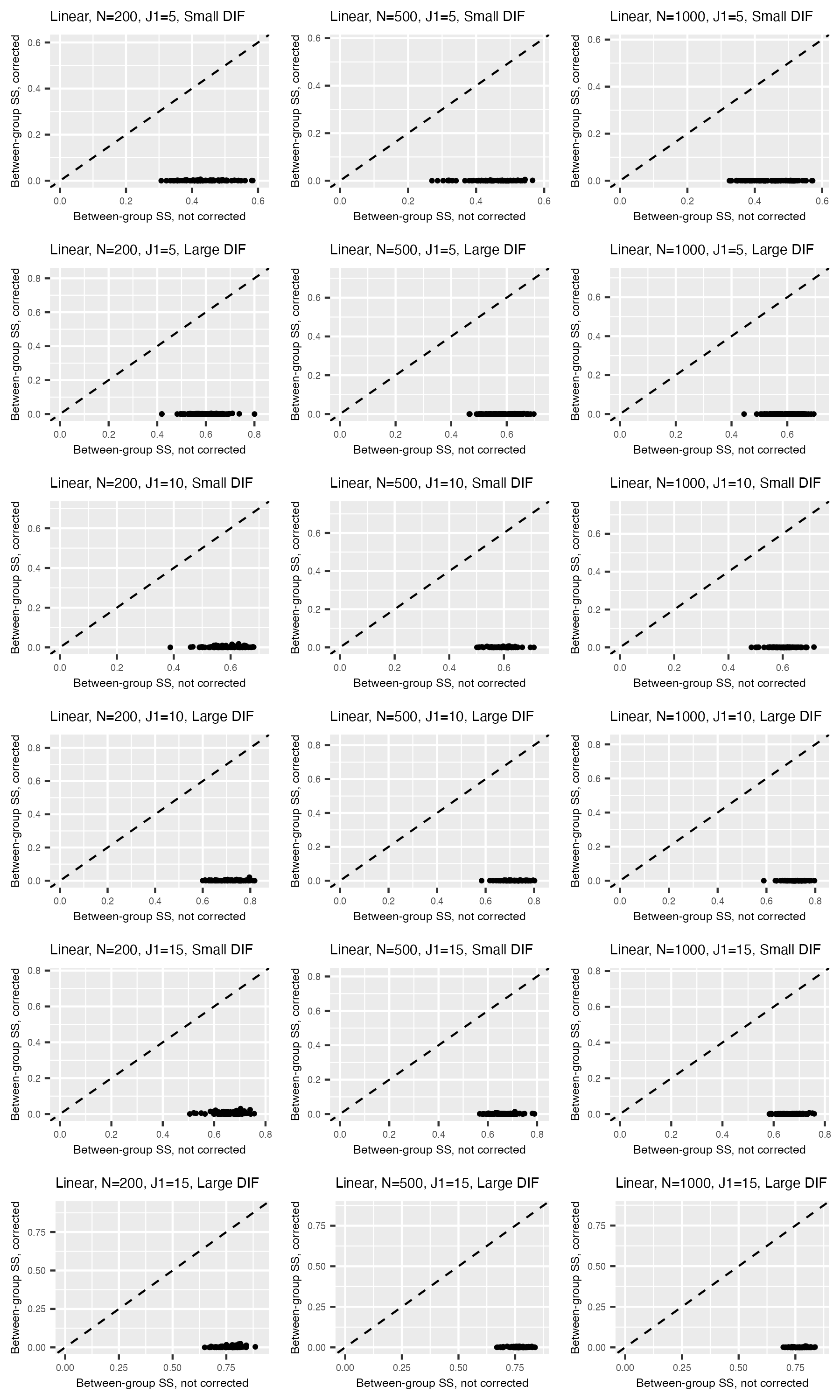}
    \caption{Between-group sum of squared bias for target trait estimation for the \emph{linear} model with \emph{non-uniform} DIF, under different simulation settings. The x-axis corresponds to the estimation without DIF correction using the DIF items; the y-axis corresponds to the DIF-corrected estimation using the DIF items.}
    \label{fig:sim_anova_linear_nonuniform}
\end{figure}

\begin{figure}
    \centering
    \includegraphics[width=0.63\linewidth]{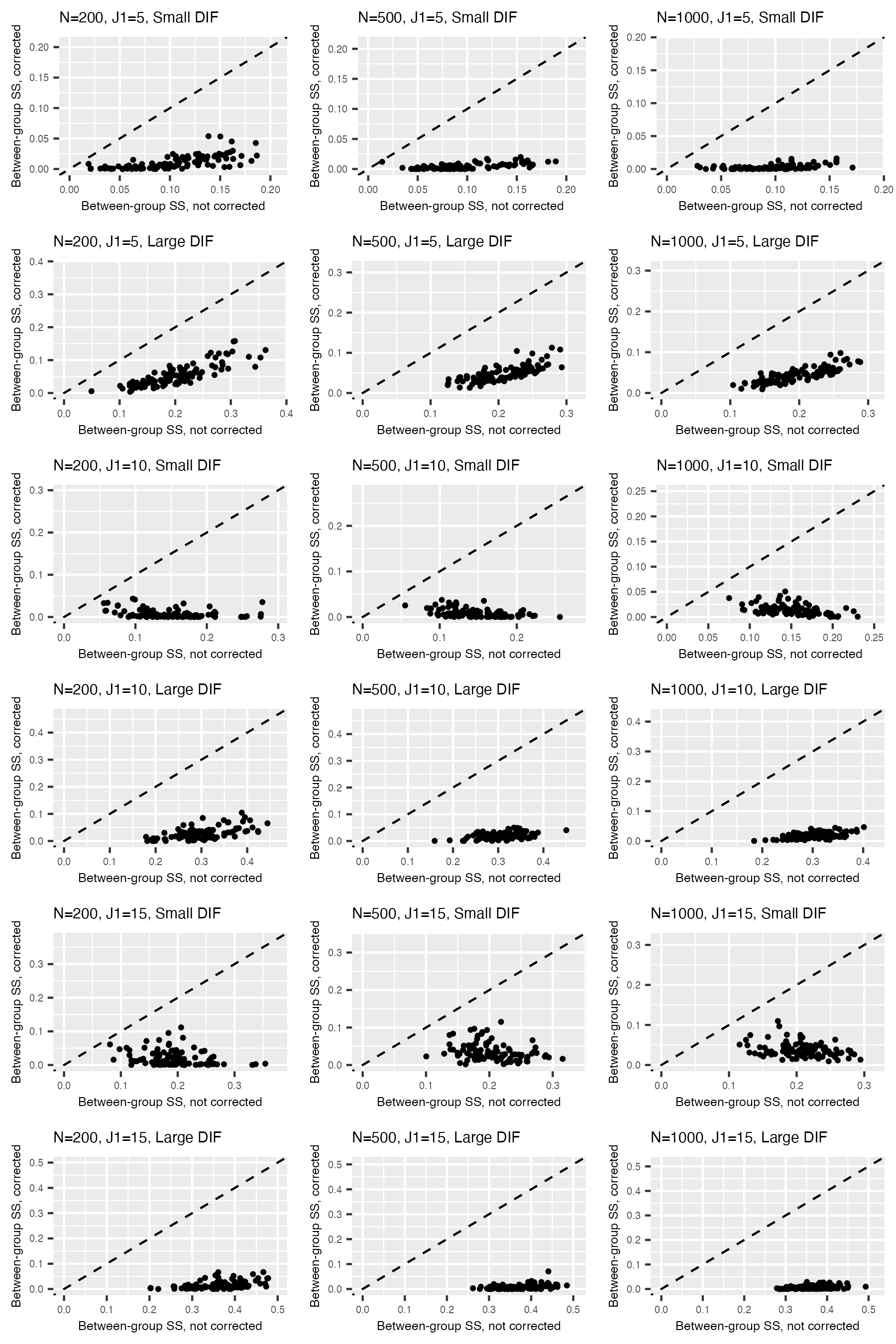}
    \caption{Between-group sum of squared bias for target trait estimation for the \emph{M2PL} model with \emph{non-uniform} DIF, under different simulation settings. The x-axis corresponds to the estimation without DIF correction using the DIF items; the y-axis corresponds to the DIF-corrected estimation using the DIF items.}
    \label{fig:sim_anova_m2pl_nonuniform}
\end{figure}

\clearpage
\subsection*{B.3. Case Study}

\begin{figure}[hbt!]
    \centering
    \includegraphics[width=0.9\textwidth]{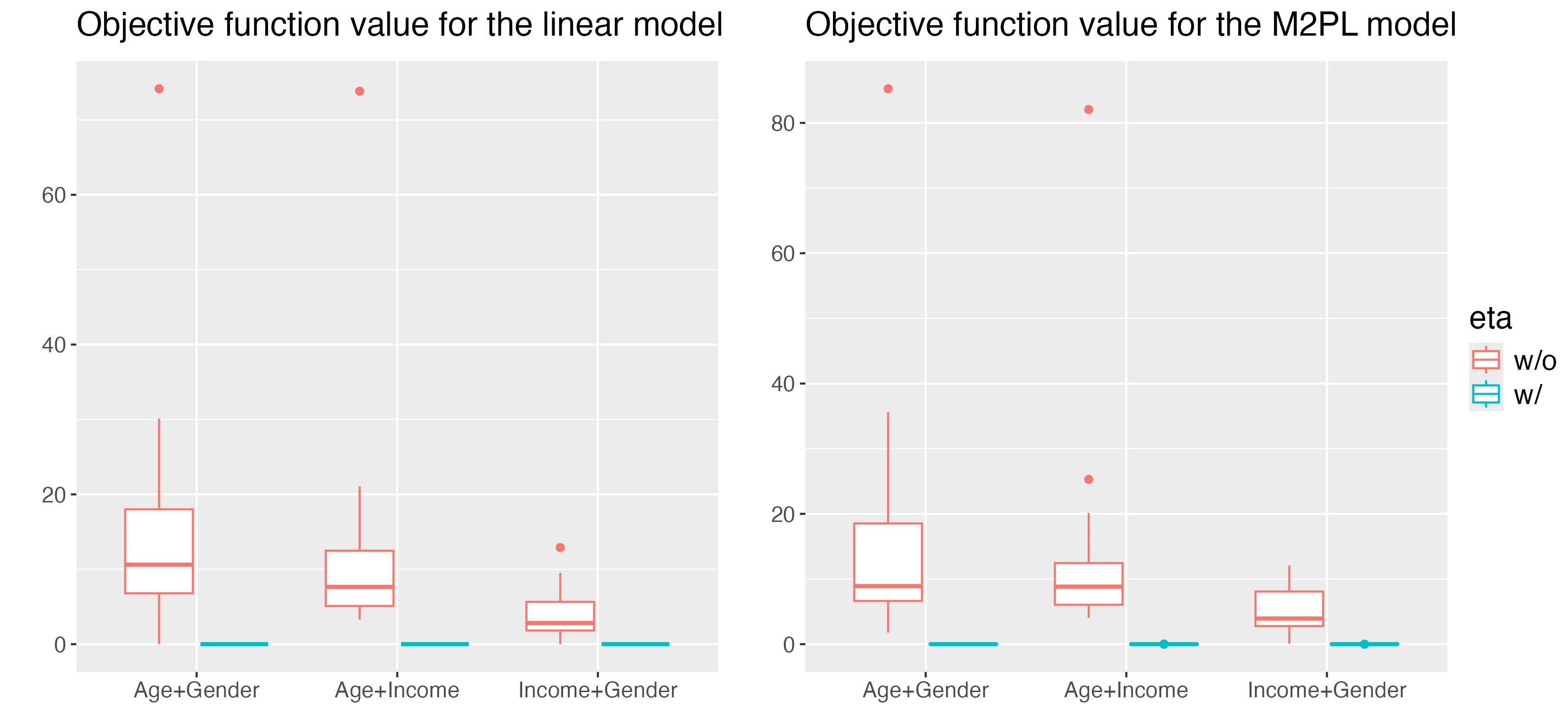}
    \caption{Comparing the objective function value with and without the nuisance trait surrogate for the linear (left) and the M2PL model (right) with two grouping variables.}
    \label{fig:obj_2Z}
\end{figure}

\vspace{\fill}\pagebreak

\end{document}